\documentclass[sigconf]{acmart}
\usepackage{pifont}
\usepackage{colortbl}
\usepackage{array}
\usepackage{multirow}
\usepackage{amsmath}
\usepackage{graphicx}
\usepackage{subcaption}
\definecolor{latencyregion}{HTML}{EAF3FF}
\definecolor{intermediateregion}{HTML}{FFF4D6}
\definecolor{runtimeregion}{HTML}{E8F6EA}

\usepackage{cleveref}

\AtBeginDocument{%
  }

\copyrightyear{2026}
\acmYear{2026}
\setcopyright{cc}
\setcctype{by}
\acmConference[SoCC '26 Vol. 2]{ACM Symposium on Cloud Computing Volume 2}{November 18--20, 2026}{Singapore, Singapore}
\acmBooktitle{ACM Symposium on Cloud Computing Volume 2 (SoCC '26 Vol. 2), November 18--20, 2026, Singapore, Singapore}
\acmDOI{10.1145/3828161.3857877}
\acmISBN{979-8-4007-2816-7/2026/11}

\begin{document}

\title{(Near-)Pareto-Optimal Synthesized OS-Level Scheduling for Serverless Systems}
\renewcommand{\shorttitle}{(Near-)Pareto-Optimal Synthesized OS-Level Scheduling for Serverless Systems}

% \author{Yuxuan Zhao$^{1}$\quad
%         Zhao Yang$^{2}$\quad
%         Weikang Weng$^{1}$\quad\\
%         Vincent François-Lavet$^{2}$\quad
%         Rob van Nieuwpoort$^{1}$\quad
%         Alexandru Uta$^{3}$}
% \affiliation{%
%   \institution{$^{1}$Leiden University\quad
%                $^{2}$VU Amsterdam\quad
%                $^{3}$DFINITY}
%   \country{}
% }
\author{Yuxuan Zhao}
\affiliation{%
  \institution{Leiden University}
  \city{Leiden}
  % \state{Texas}
  \country{Netherlands}
  }
\email{y.zhao@liacs.leidenuniv.nl}

\author{Zhao Yang}
\affiliation{%
  \institution{VU Amsterdam}
  \city{Amsterdam}
  % \state{Texas}
  \country{Netherlands}
  }
\email{z.yang3@vu.nl}

\author{Weikang Weng}
\affiliation{%
  \institution{Leiden University}
  \city{Leiden}
  % \state{Texas}
  \country{Netherlands}
  }
\email{w.weng@liacs.leidenuniv.nl}

\author{Vincent François-Lavet}
\affiliation{%
  \institution{VU Amsterdam}
  \city{Amsterdam}
  % \state{Texas}
  \country{Netherlands}
  }
\email{vincent.francoislavet@vu.nl}

\author{Rob van Nieuwpoort}
\affiliation{%
  \institution{Leiden University}
  \city{Leiden}
  % \state{Texas}
  \country{Netherlands}
  }
\email{r.v.van.nieuwpoort@liacs.leidenuniv.nl}

\author{Alexandru Uta}
\affiliation{%
  \institution{DFINITY}
  \city{Zurich}
  % \state{Texas}
  \country{Switzerland}
  }
\email{alexandru.uta@gmail.com}

\renewcommand{\shortauthors}{Yuxuan Zhao et al.}
\renewcommand{\authors}{Yuxuan Zhao, Zhao Yang, Weikang Weng, Vincent François-Lavet, Rob van Nieuwpoort, and Alexandru Uta}




\begin{abstract}
Serverless cloud platforms typically rely on operating-system schedulers such as the Completely Fair Scheduler (CFS) and Earliest Eligible Virtual Deadline First (EEVDF) to schedule function executions across CPU cores. Prior work has shown that existing schedulers either minimize task latency at the expense of runtime efficiency or improve runtime efficiency while increasing latency. Although hybrid approaches partially mitigate this trade-off, each scheduler remains confined to a narrow region of the latency--runtime trade-off space, leaving much of the design space unexplored.

We present a class of synthesized schedulers that expose configurable latency--runtime trade-offs. Rather than hand-designing heuristics, we use reinforcement learning to synthesize policies for core assignment, preemption, and load balancing under different reward objectives. We implement the synthesized schedulers using Linux \texttt{sched\_ext}, train them on representative serverless workloads, and evaluate them on a multi-core server. Synthesized schedulers are competitive with both general-purpose Linux schedulers (CFS, EEVDF, and FIFO) and state-of-the-art serverless-aware schedulers. More importantly, they can be tuned at deployment time to span a near-continuous set of latency--runtime operating points. Several of these operating points reduce both latency and runtime cost, improving profitability for both cloud users and providers.
\end{abstract}

\maketitle

\vspace{-0.15cm}
\section{Introduction}

Serverless computing has emerged as a cloud paradigm in which users run functions without managing the servers themselves. Most serverless platforms are built on Function-as-a-Service (FaaS), which executes functions in response to events and automatically scales resources with workload demand~\cite{van2018serverless, shahrad2020serverless, hellerstein2018serverless, jonas2019cloud, roberts2017serverless, zhao2022tiny}. This abstraction simplifies application deployment and shifts the responsibility for efficient execution to the cloud provider. Because users are charged under a pay-as-you-go model, they only pay for the runtime of function invocations~\cite{jonas2019cloud, eivy2017wary, elgamal2018costless, AWSLambdaPricing}. Once functions are running on a worker node, their latency, runtime, and billing cost are strongly affected by how they are scheduled on the server. Thus, node-level scheduling decisions directly influence both user-perceived performance and provider-side efficiency.

OS-level scheduling practice has not kept pace with this shift: node-level scheduling still relies on general-purpose OS schedulers that were not designed for serverless execution~\cite{fu2022sfs,zhao2024serverless,fu2024alps}, and they can become a performance bottleneck. Traditional Linux schedulers, such as the Completely Fair Scheduler (CFS)~\cite{cfs} and the Earliest Eligible Virtual Deadline First Scheduler (EEVDF)~\cite{eevdf}, aim to balance fairness, latency, and throughput across diverse workloads~\cite{lozi2016linux, koller2017will}. However, serverless workloads differ from the long-running processes for which these schedulers are typically optimized: they are characterized by high concurrency, bursty invocations, and short-lived tasks that often complete in milliseconds to seconds~\cite{shahrad2020serverless, nasiri2026production, joosen2023does}. 

\begin{figure}[tp]
    \centering
    \includegraphics[width=0.88\linewidth]{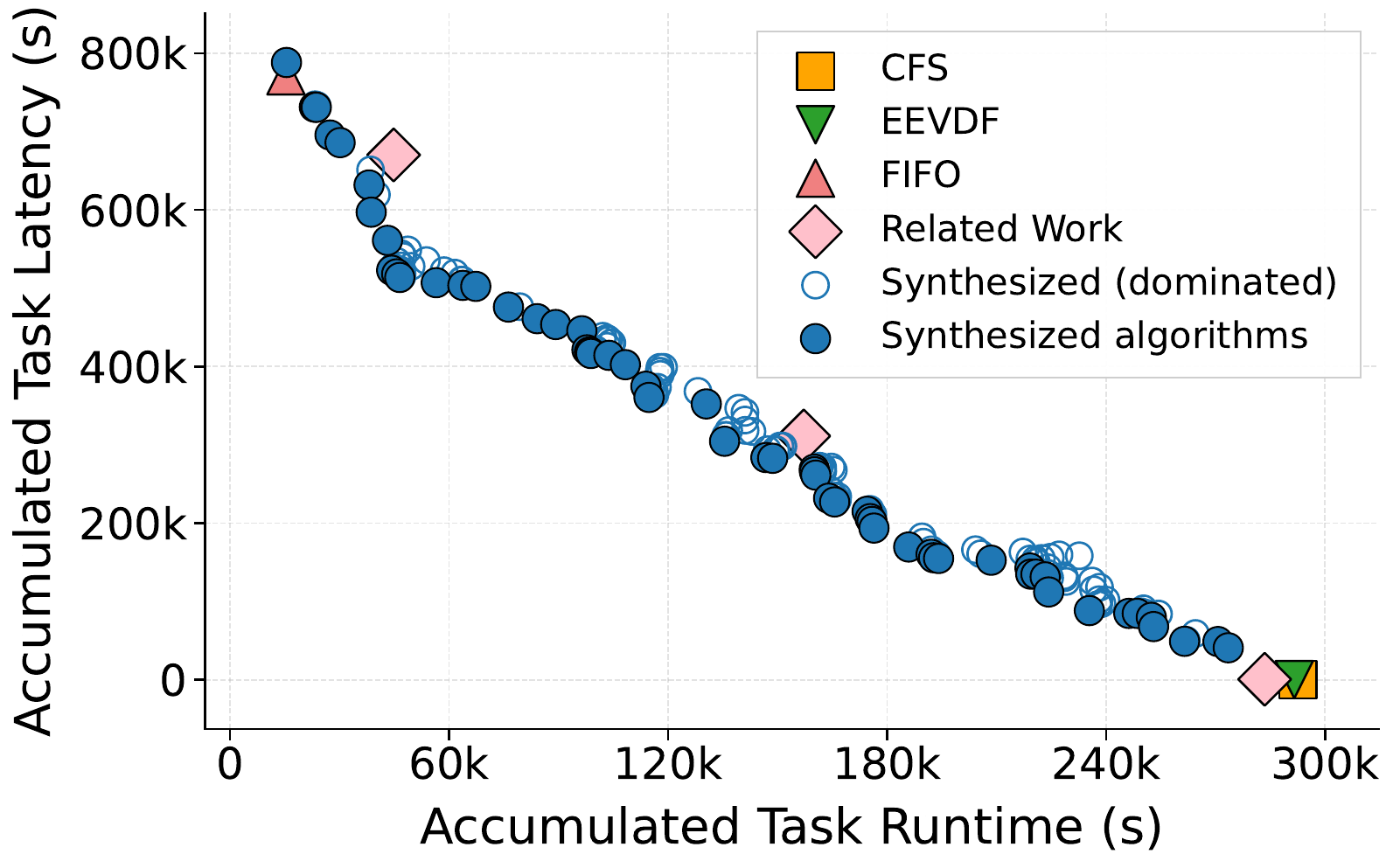}
    \vspace{-0.4cm}
    \caption{Latency--runtime trade-off. Accumulated task latency and accumulated task runtime across a workload extracted from Microsoft Azure trace. CFS and EEVDF achieve low latency with high runtime, while FIFO reduces runtime at the cost of higher latency. Our synthesized schedulers trace out an empirical Pareto frontier.}
    
    \label{fig:pareto_scatter_middle_dense_seeds}
    \vspace*{-0.5cm}
\end{figure}

Under such conditions, fairness mechanisms across tasks based on historical execution can be less effective, because many functions finish before the scheduler can accumulate meaningful execution history~\cite{koller2017will}. Moreover, serverless workers are often highly overcommitted, causing frequent context switches, preemptions, and contention for shared hardware resources such as CPU cores, memory bandwidth, and last-level cache~\cite{isstaif2023towards, merkel2008memory, suresh2021servermore, gajanin2025performance}. These effects can increase the task runtime and therefore billing cost~\cite{isstaif2023towards, zhao2024serverless, fu2022sfs, lin2026demystifying}, while also degrading user-perceived latency~\cite{tian2022owl, li2023golgi}. To address these latency and contention problems, prior low-latency scheduling systems introduce mechanisms such as centralized dispatch, preemption, and core-granular load balancing~\cite{kaffes2019shinjuku, iyer2023achieving, kaffes2019centralized}. These policies, however, are fixed: each offers a single scheduling behavior, whereas serverless applications are highly heterogeneous---spanning latency-critical to cost-sensitive requirements~\cite{shahrad2020serverless, joosen2023does, nasiri2026production} and shifting across execution phases---that no single predetermined policy can satisfy at once.

These limitations motivate reasoning about the full space of scheduling behaviors rather than any single fixed policy. In practice, cloud servers handling serverless workloads are provisioned for long periods of time, continuously running functions for spans such as weeks, months, or longer. To increase profits, providers keep these machines continuously at a high utilization level~\cite{agache2020firecracker,brooker2023demand}, making the OS scheduler one of the central parts of the cloud stack. When servers are continuously running tasks, there are two important high-level metrics that can be used to reason about performance: function latency and function runtime. Both are user-facing, with the former having a high impact in deadline-constrained tasks, while the latter is directly proportional to billed cost~\cite{AWSLambdaPricing}. Moreover, runtime is extremely important for providers as well: optimizing for short runtimes leaves room for improved overall throughput and cost consolidation. Therefore, when considering a period of time in the life of a cloud server, \emph{accumulated latency} and \emph{accumulated runtime} paint a clear picture of the OS scheduler performance, e.g., as presented in Figure~\ref{fig:pareto_scatter_middle_dense_seeds}. Existing policies such as CFS, EEVDF or FIFO occupy distinct operating points within the broader latency--runtime trade-off space.   

This raises the central question of this paper: what operating points are available in the latency--runtime trade-off space for serverless OS scheduling, and can OS schedulers be systematically tuned to move toward different objectives? Instead of committing to a single fixed objective, we explore an empirical Pareto frontier between accumulated task latency and accumulated task runtime. Throughout this paper, we use the term Pareto in its empirical sense: an operating point lies on the empirical Pareto frontier if no other configuration we measured improves on it in both accumulated latency and accumulated runtime. This perspective is important because CFS, EEVDF, and FIFO occupy different regions of the trade-off space: CFS/EEVDF reduce waiting time but incur longer runtime (hence higher user cost), while FIFO preserves runtime efficiency but increases user-perceived latency.

Learning-based scheduling provides a way to synthesize policies for different points in this latency--runtime trade-off space. Reinforcement learning (RL) is a good fit for this setting because scheduling is a sequential decision-making problem in which actions can be evaluated through feedback from the environment~\cite{sutton1998reinforcement, mao2019learning}. Recent studies have applied RL to serverless environments, mainly focusing on fleet optimization and resource management~\cite{yu2021faasrank, qiu2022simppo}. However,  these approaches treat each worker node as a black box. After a function is assigned to a server, its execution is governed by the local OS scheduler. As a result, cluster-level placement alone cannot address local effects such as context switching, queue imbalances, and preemption behavior. These effects directly influence function latency, runtime, and therefore billing cost.

To explore the empirical latency--runtime Pareto space (Figure~\ref{fig:pareto_scatter_middle_dense_seeds}), we present an architecture for OS scheduler synthesis. It uses RL algorithms to drive low-level OS scheduling mechanisms under different reward objectives. Applying RL at the OS scheduling layer, however, requires decisions at much finer timescales than typical learning-based resource managers can support. We therefore separate scheduling control across two timescales, amortizing inference through batched task placement while adapting system-level preemption and rebalancing less frequently. Our synthesized schedulers use \texttt{sched\_ext}~\cite{schedext} to achieve efficient user-space scheduling. Schedulers use two RL policies: the high-frequency policy handles CPU core assignment for arriving serverless tasks, while the low-frequency policy observes global system state, such as queue lengths, to adjust the preemption time slice and to perform load balancing. By observing the runtime state of the execution environment, the two policies learn to optimize end-to-end objectives such as accumulated task latency and accumulated task runtime. In this article, we make the following contributions:
\begin{itemize}
    \item We design and implement an architecture to synthesize OS-level schedulers for serverless functions. Based on RL-driven decisions, the synthesized schedulers couple high-frequency task-dispatch decisions with a lower-frequency controller that adjusts preemption and load rebalancing.
    \item We synthesize and deploy schedulers on multi-core systems using a modern extensible scheduling framework (\texttt{sched\_ext}) and evaluate them with real-world FaaS workload traces. Our synthesized schedulers are competitive with both general-purpose (CFS, EEVDF, FIFO) and serverless-aware (Hybrid~\cite{zhao2024serverless}, ALPS~\cite{fu2024alps}, SFS~\cite{fu2022sfs}) schedulers.
    \item We explore and characterize the empirical latency--runtime Pareto space of serverless OS scheduling. We show that synthesized schedulers offer trade-off points beyond what existing schedulers can offer, leading to increased profitability for both clients and cloud providers.
\end{itemize}

\vspace{-0.15cm}
\section{Background}
In this section, we introduce background knowledge for OS-Level scheduling for serverless, reinforcement learning, and two reinforcement learning methods used in this article.

\vspace{-0.15cm}
\subsection{OS-Level Scheduling for Serverless}
A serverless worker executes many concurrent function invocations on a multi-core machine. Once functions are assigned to the worker, their latency and runtime are determined not only by the function itself but also by the OS-level scheduler. In this work, we define task latency as the elapsed time from arrival to first execution, and task runtime as the elapsed time from first execution to its completion. Under this definition, runtime can increase when a task is preempted, requeued, or migrated after it has started. The scheduler therefore affects both when a function first executes and how long it takes to complete after starting. 

The system must make online decisions under limited information. In particular, task runtimes are unknown a priori and may vary significantly across invocations, preventing accurate workload prediction. The scheduler observes only coarse-grained system state, such as per-core queue lengths/occupancy, rather than detailed per-task remaining processing time. Arriving tasks are independent and must be assigned without knowledge of future arrivals. Additionally, once dispatched, tasks may be subject to system constraints such as non-negligible context switching or migration overheads. These factors jointly make the scheduling problem challenging, as decisions must balance load distribution and latency with only partial information about the workload or future arrivals.

A scheduler decides where arriving functions are queued, when running functions should be preempted, and whether queued work should be rebalanced across cores. These decisions create a latency--runtime trade-off. Giving arriving functions earlier CPU access can reduce latency, but frequent preemption may increase runtime because a preempted task's measured runtime continues to grow while it waits to resume execution, in addition to context-switch overhead and reduced cache locality. In contrast, run-to-completion execution can preserve runtime efficiency, but may cause head-of-line blocking when long functions occupy per-core queues. Similarly, moving tasks across cores can reduce queue imbalance, but excessive migration may introduce overhead. These mechanisms form the core scheduling space explored in this work.

\vspace{-0.15cm}
\subsection{Reinforcement Learning}
Reinforcement learning (RL)~\cite{sutton1998reinforcement,vincent2018introduction} addresses sequential decision-making problems that are typically formalized as a Markov Decision Process (MDP). An MDP is defined by a tuple $(\mathcal{S}, \mathcal{A}, P, R, \gamma)$, where $\mathcal{S}$ is the state space, $\mathcal{A}$ is the action space, $P(s'|s,a)$ is the state transition probability, $R(s,a)$ is the reward function, and $\gamma \in [0,1)$ is the discount factor. At each time step $t$, an agent observes state $s_t \in \mathcal{S}$, selects action $a_t \in \mathcal{A}$ according to a policy $\pi(a_t|s_t)$, receives reward $r_t = R(s_t, a_t)$, and transitions to the next state $s_{t+1} \sim P(\cdot|s_t, a_t)$.
The goal of RL is to find an optimal policy $\pi^*$ that maximizes the expected cumulative discounted reward:
\begin{equation}
    \pi^* = \arg\max_{\pi} \mathbb{E}_{\pi} \left[ \sum_{t=0}^{\infty} \gamma^t r_t \right]
\end{equation}
One approach to learning the optimal policy is through value-based methods, which first learn a state-action value function (Q-function) $Q^\pi(s,a)$ that represents the expected return when taking action $a$ in state $s$ and following policy $\pi$ thereafter:
\begin{equation}
    Q^\pi(s,a) = \mathbb{E}_{\pi} \left[ \sum_{k=0}^{\infty} \gamma^k r_{t+k} \mid s_t = s, a_t = a \right]
\end{equation}
Once the optimal Q-function $Q^*$ is learned, the optimal policy can be derived by acting greedily with respect to the Q-values: $\pi^*(s) = \arg\max_a Q^*(s,a)$.

\vspace{-0.15cm}
\subsection{DQN and Dueling DQN}
Deep Q-Networks (DQN)~\cite{mnih2013playing} use a deep neural network to estimate action values, allowing RL to scale to complex, high-dimensional environments. The network is trained with temporal-difference learning, minimizing the difference between its current prediction and a target that combines the immediate reward with the discounted maximum predicted value of the next state. Two mechanisms stabilize training: experience replay stores past transitions in a buffer and trains on randomly sampled mini-batches, breaking the correlation between consecutive steps; and a target network, a frozen copy of the network synchronized at fixed intervals, generates the targets, preventing them from shifting too rapidly and confusing the optimization process.

Standard DQN learns one Q-value per action directly, without separating a state's baseline value from the relative advantage of each action, which is wasteful when many actions have similar values. Dueling DQN~\cite{wang2016dueling} instead splits the network after a shared feature-extraction layer into two heads: a state-value stream estimating the overall benefit of the current state, and an advantage stream estimating the relative merit of each action. The two are merged into Q-values, with the advantage stream centered by subtracting its mean to ensure that the system can uniquely identify the individual contributions of the state value and action advantages.

\vspace{-0.15cm}
\section{System Model}
\label{section:motivation}

In this section, we explain the main design choices behind our synthesized scheduler before formalizing the scheduling problem. We start from a simple scheduler that assigns functions to per-core FIFO queues and show why two scheduling mechanisms, preemption and work shuffling, are needed for serverless workloads (\Cref{subsec:preemption_workshuffling}). We then justify the choice of Q-learning as the RL algorithm for our synthesized scheduler (\Cref{subsec:qlearning}). Next, we explain why a global queue is necessary to expose system-wide state and support batched decisions in our reinforcement learning architecture (\Cref{subsec:global_queue}). Finally, we define the scheduling problem and its state/action spaces (\Cref{subsec:problem_statement}).

\vspace{-0.15cm}
\subsection{Scheduling Mechanisms}\label{subsec:preemption_workshuffling}

In serverless clouds, the underlying servers run for weeks or months and are kept busy at high utilization~\cite{agache2020firecracker,brooker2023demand}. The OS scheduler decides how the incoming functions are placed and run on servers, hence the scheduler has a direct effect on two metrics that both users and cloud providers deem important. One is latency, which has impact on deadline-constrained functions. The other is runtime, which decides how much the user is billed under a pay-as-you-go pricing model~\cite{AWSLambdaPricing}. For the provider, runtime also influences how much capacity there is to improve the overall throughput and achieve cost consolidation. We therefore measure a scheduler by its \emph{accumulated latency} and \emph{accumulated runtime} over a workload, and use these two metrics to motivate the system model below.

\begin{figure}[t]
    \centering
    % \vspace{-0.3cm}
    \begin{subfigure}[b]{0.48\linewidth}
        \centering
        \includegraphics[width=\linewidth]{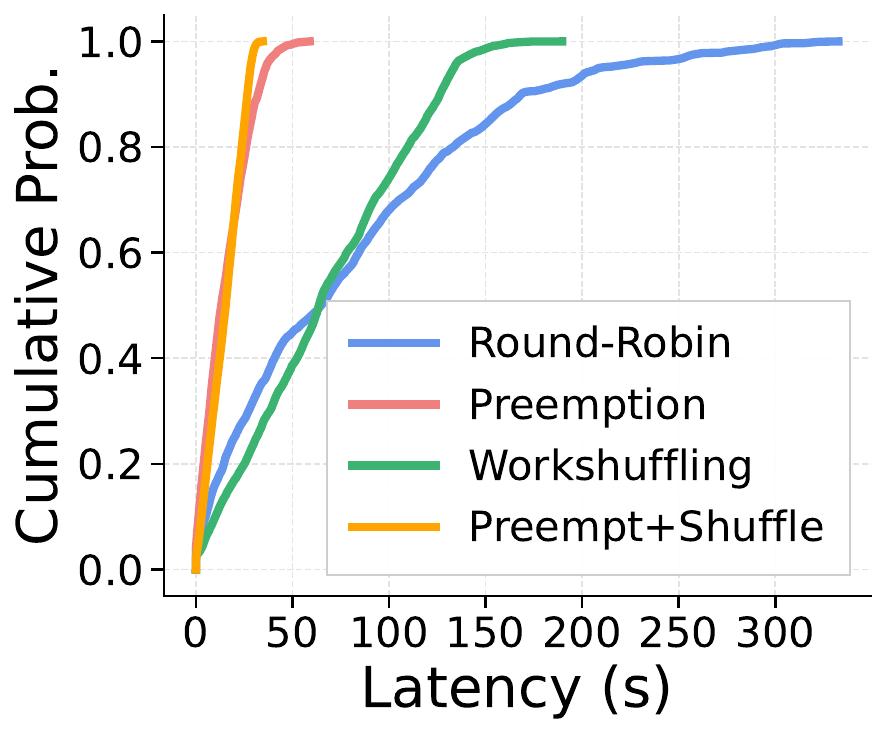}
        \vspace{-0.55cm}
        \caption{CDFs of response latency.}
        \label{fig:response-time}
    \end{subfigure}
    \begin{subfigure}[b]{0.48\linewidth}
        \centering
        \includegraphics[width=\linewidth]{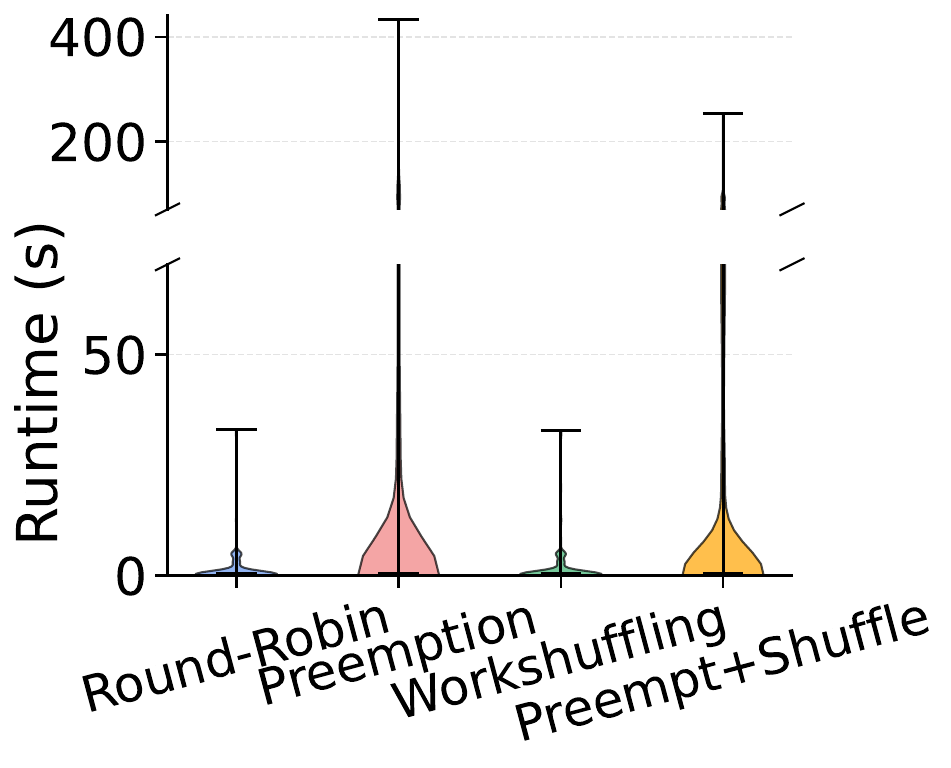}
        \vspace{-0.55cm}
        \caption{Runtime distributions.}
        \label{fig:execution-time}
    \end{subfigure}
    \vspace{-0.3cm}
    
    \caption{CDFs of response latency and runtime distributions with a workload under Round Robin, Preemption, Work shuffling on 50 CPU cores.}
    \label{fig:combined-metrics}
    \vspace*{-0.5cm}
\end{figure}

When building our synthesized schedulers, we build on mechanisms used by existing schedulers: global/per-core queues, preemption, and work shuffling. However, we use RL policies to steer these mechanisms. To showcase the integration of preemption and work shuffling into our synthesized scheduler, we first conduct a baseline experiment. We evaluate a workload of 9,463 serverless functions on 50 CPU cores and compare the latency and runtime distributions under four scheduling mechanisms, as follows:

\textbf{Round-Robin. }
In \textit{round-robin}, tasks are assigned to per-core queues in a round-robin manner and are executed in a First-In-First-Out (FIFO) order. Neither preemption nor work shuffling is enabled. This scheduler serves as our baseline.

\textbf{Preemption. }
Tasks are initially assigned in a round-robin manner and executed in FIFO order. A fixed time slice of 1,000\,ms is applied. Any task exceeding this runtime quota is preempted and requeued at the end of its per-core queue.

\textbf{Work Shuffling}
is a load balance scheduling mechanism. Tasks are initially assigned in a round-robin manner and executed in FIFO order. When the most heavily loaded core has more than 20 queued tasks, the scheduler migrates half of the queue-length difference to the least-loaded core. To avoid migration thrashing (i.e., the ping-pong effect), shuffling is triggered only when the queue length difference exceeds 25\% of the maximum queue length.

Figures~\ref{fig:response-time} and~\ref{fig:execution-time} show the latency and runtime distributions, respectively. As shown, \textit{Preemption} mechanism reduces latency for shorter functions by mitigating head-of-line blocking. However, this benefit comes at the cost of longer runtime, defined as the duration from first time run to final completion. This is because preempted tasks may incur additional context-switch overhead and waiting time after being requeued. By contrast, \textit{Work Shuffling} targets a different bottleneck. Rather than interrupting running tasks, it mitigates spatial queue imbalance across CPU cores. As shown in Table~\ref{tab:compare_mechanisms}, work shuffling mechanism reduces long-tail effect compared to other scheduling mechanisms. This improvement occurs because the workload is less likely to be bottlenecked by a few congested per-core queues, thereby reducing tail delays that would otherwise prolong the entire workload's completion time.

In summary, \textit{Preemption} and \textit{Work Shuffling} improve different aspects of serverless scheduling performance. \textit{Preemption} reduces short-function latency by mitigating head-of-line blocking, but it may increase runtime due to requeueing and context-switch overhead. \textit{Work Shuffling} reduces the long-tail effect by mitigating spatial queue imbalance across CPU cores.

\vspace{-0.15cm}
\subsection{Scheduler Global Queue}\label{subsec:global_queue}

We motivate the use of a global queue in our reinforcement learning architecture. The global queue provides two key benefits.

\textbf{1. Global State Visibility.} Per-core queues only expose local load information. In contrast, a global queue allows the RL agent to observe the system-wide arrival pattern and the distribution of pending work across CPU core queues. This visibility is important for making spatial load-balancing decisions and for penalizing queue imbalance during learning.

\textbf{2. Temporal Batching and Burst Awareness.} Serverless workloads are often bursty, and dispatching each task immediately upon arrival would force the scheduler to make decisions at a very high frequency. Instead, the global queue buffers incoming tasks over short time windows, such as 20\,ms, and dispatches them as a batch. This reduces the frequency of RL inference and allows the scheduler to estimate the arrival intensity of incoming traffic. For example, if a batch contains 100 tasks, the scheduler can infer that the system is experiencing a burst and may apply more aggressive load balancing or preemption. Conversely, if a batch contains only a few tasks, the scheduler may choose a more conservative policy.

\begin{table}[t]
    % \vspace{-0.4cm}s
    \caption{Comparison of the time required to complete 99\% of the tasks under four scheduling mechanisms.}
    \vspace{-0.4cm}
    \centering
    \footnotesize
    \setlength{\tabcolsep}{4pt}
    \begin{tabular*}{0.9\columnwidth}{@{\extracolsep{\fill}}l c@{}}
        \toprule
        \textbf{Mechanisms} & \textbf{Time to Complete 99\% of Tasks (s)} \\
        \midrule
        Round-Robin & 388 \\
        Preemption & 294 \\
        Work Shuffling & 268 \\
        Preempt+Shuffle & 254 \\
        \bottomrule
    \end{tabular*}
    \label{tab:compare_mechanisms}
    \vspace{-0.5cm}
\end{table}

\vspace{-0.15cm}
\subsection{Q-learning}\label{subsec:qlearning}

The observation above suggests that \textit{Preemption} and \textit{Work Shuffling} are not universally beneficial. Their effectiveness depends on the dynamic system state, including workload burstiness, queue imbalances, and workload characteristics. This calls for an adaptive policy that decides when to apply each mechanism. However, designing an adaptive scheduler that jointly adjusts preemption and shuffling from arrival rate and queue imbalance, and stays valid as load shifts, is non-trivial. For example, ten tasks that have not started and ten waiting to resume give the same arrival rate and queue length, but need different handling. Preemption also has a delayed cost: the interrupted task keeps accruing runtime while it waits, and how long it waits depends on where it was placed.

We therefore formulate this adaptive scheduling problem as a reinforcement learning problem, in which the agent observes burst intensity and system-wide state. Because each decision changes the subsequent queueing state and reward accumulates over a sequence of decisions, the policy is trained on the cumulative outcome rather than on the immediate per-decision improvement that a greedy heuristic targets. Concretely, the scheduler selects a destination CPU core for each incoming task, chooses a preemption time slice from a finite set, and selects destinations for tasks to be shuffled. All of these actions are discrete and the action space is small, which makes value-based methods a natural and sample-efficient fit. To handle the high-dimensional state space and exploit the generalization of neural networks, we use deep Q-learning, specifically DQN~\cite{mnih2013playing} and Dueling DQN~\cite{wang2016dueling}. Policy-gradient methods such as DDPG~\cite{lillicrap2016continuous}, TD3~\cite{fujimoto2018addressing}, and SAC~\cite{haarnoja2018soft}, by contrast, target continuous action spaces and would require modification to fit our discrete formulation. Discrete policy-gradient methods also apply to our setting, and we leave a comparison for future work.

In practice, a scheduling decision spans core assignment, the preemption time slice, and load rebalancing. Rather than encoding all of these into a single large discrete action space, we decompose control into two value-based agents: a high-frequency agent that selects destination cores for spatial placement, and a low-frequency agent that selects the preemption time slice for temporal control. Load rebalancing reuses the placement agent to choose migration targets. Because each agent operates over a much smaller action space than their combination, the two spaces stay compact, which in turn stabilizes learning and improves sample efficiency.

\vspace{-0.15cm}
\subsection{Objectives}\label{subsec:problem_statement}

We model serverless function scheduling as an MDP. The RL agent observes the current execution state and incoming function invocations, then selects scheduling actions, including the target CPU core, preemption time slice, and shuffle destination. The resulting state and action spaces are summarized in Table~\ref{tab:state-action}.

\begin{table}[t]
    \caption{State and action spaces of the RL agent (in parentheses, the number of discrete choices for each action).}
    \vspace{-0.4cm}
    \centering
    \begin{tabular}{p{8cm}}
        \hline
        \rowcolor[gray]{0.9}\centering\textbf{State Space}\tabularnewline
        Number of started tasks, number of unstarted tasks, number of finished tasks, per-core queue lengths, accumulated latency, accumulated runtime, accumulated preemption count \\
        
        \rowcolor[gray]{0.9}\centering\textbf{Action Space}\tabularnewline
        Destination CPU core for incoming invocations (50 choices DDQN~\cite{wang2016dueling}), preemption time slice (5 choices, DQN~\cite{mnih2013playing}), destination CPU cores for rebalanced tasks (50 choices, DDQN~\cite{wang2016dueling}) \\
        \hline
    \end{tabular}
    \label{tab:state-action}
    \vspace{-0.5cm}
\end{table}

Since serverless platforms continuously execute function invocations, scheduling performance can be characterized by two key metrics: latency and runtime. Latency is critical for deadline-constrained user-facing workloads, whereas runtime is directly related to billed cost under the pay-as-you-go pricing model. The objective of the RL agent is to maximize cumulative reward over workload execution while optimizing \emph{accumulated task latency} and \emph{accumulated task runtime}. We aim to synthesize two classes of schedulers, each optimized for a different reward objective:

\textbf{Latency-oriented schedulers}  optimize accumulated task latency, defined as the sum of the time from function arrival to first execution across all tasks. This objective is important for user-facing services, where low latency directly affects the user experience. Prior work reports that a one-second slowdown can reduce customer satisfaction by 16\%~\cite{speed-matters,kleppmann2017designing}.

\textbf{Runtime-oriented schedulers} optimize accumulated task runtime, defined as the sum of the time from first execution to completion across all tasks. This objective is important because users are charged under a pay-as-you-go pricing model, where they pay for the runtime of their functions~\cite{jonas2019cloud, eivy2017wary, elgamal2018costless}.

\vspace{-0.15cm}
\section{Synthesizing Schedulers}
This section presents the design and implementation of our scheduler-synthesis architecture. We first give an overview of the  architecture, then describe the reward designs that steer a policy toward low latency or low runtime. We then show how we train convergent policies, how a trained policy is tuned to explore the latency--runtime trade-off, and how the scheduler is implemented on \texttt{sched\_ext}.

\vspace{-0.15cm}
\subsection{Architecture Overview}
Driven by the motivation in~\Cref{section:motivation}, we propose a scheduler-synthesis architecture based on reinforcement learning.

\textbf{System Overview.} Our schedulers follow a centralized, single-dispatcher model. The kernel-side BPF backend enqueues every runnable task into a global shared queue (a ring buffer spanning all CPUs), which serves as the system-wide global queue. Tasks arriving at the system are initially buffered in this global queue. A single user-space scheduler thread drains the global queue, decides placement, and dispatches each task to the tail of a specific per-core queue; each worker core owns one such per-core queue, within which tasks are processed in a FIFO manner. To prevent head-of-line blocking, a time-slice preemption mechanism is applied. If a task runs out of its allocated execution quota before completion, it is immediately preempted and requeued at its per-core queue's tail.

\begin{figure}[t]
    \vspace{-0.2cm}
    \centering
    \includegraphics[width=0.45\textwidth]{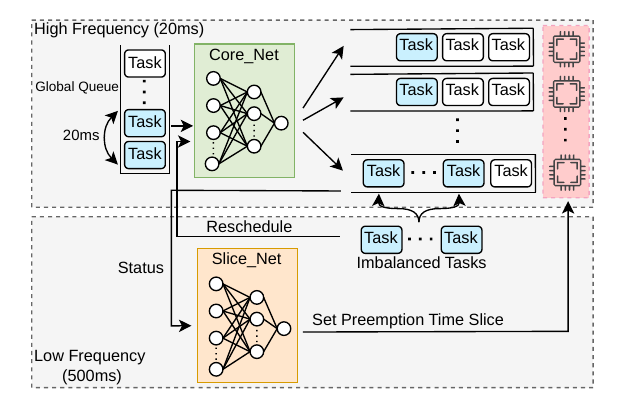}
    \vspace{-0.5cm}
    \caption{Architecture of synthesized schedulers. The high-frequency controller batches arriving tasks and assigns them to per-core queues every 20\,ms. The low-frequency controller adjusts the preemption time slice and performs load rebalancing every 500\,ms.}
    \label{fig:RL Scheduler}
    \vspace{-0.5cm}
\end{figure}

The synthesized scheduler decides both where a task runs (its spatial placement) and when it runs (its timing), so the two are coordinated rather than handled separately. The scheduler contains two learned policies that operate at different frequencies: the high-frequency policy $\pi_{\text{HF}}$, realized by Core\_Net (a Dueling DQN), and the low-frequency policy $\pi_{\text{LF}}$, realized by Slice\_Net (a DQN). The high-frequency policy handles spatial placement every 20\,ms: at each dispatch epoch, it collects the tasks accumulated in the global queue and queries $\pi_{\text{HF}}$ to assign each invocation to a worker CPU core. The low-frequency policy performs system-level tuning and rebalancing every 500\,ms: it queries $\pi_{\text{LF}}$ to adapt the global preemption time slice applied to tasks. In the same control step, it also evaluates the per-core load distribution and, when severe imbalance is detected, triggers work shuffling by migrating queued tasks from congested cores to less loaded ones. The complete architecture of our synthesized schedulers is illustrated in Figure~\ref{fig:RL Scheduler}.

\textbf{High-Frequency Policy (Spatial placement).} Operating at a fine-grained interval (e.g., 20\,ms), the high-frequency policy acts as the system's spatial task dispatcher. Instead of scheduling tasks individually upon arrival, it accumulates a batch of incoming tasks within this short window and performs a single batched inference. The policy $\pi_{\text{HF}}$ observes the current system state $s$ (e.g., per-core queue lengths, accumulated preemption count) and the number of batched tasks $n$, and outputs a vector of destination cores $\mathbf{c} = \pi_{\text{HF}}(s,n)$, where $\mathbf{c}=(c_1,\ldots,c_n)$ and $c_i$ is the worker core selected for the $i$-th task. The scheduler then dispatches each task to its per-core queue. By fixing the batching interval but letting the batch size $n$ vary with the arrival rate, the agent observes bursty workloads' instantaneous density and learns adaptive placement.

\textbf{Low-Frequency Policy (Temporal Tuning \& Rebalancing).} Operating at a coarser interval (e.g., 500\,ms), the low-frequency policy performs system-level temporal adjustment and spatial rebalancing. The policy $\pi_{\text{LF}}$ observes the global system state and selects a preemption time slice. On the worker cores, tasks are bound by this slice: once a task exhausts its quota, it is preempted. At the same interval, the low-frequency policy also evaluates the load distribution and triggers RL-driven work shuffling, migrating tasks from congested queues to idle ones to resolve spatial imbalance.

The \textbf{RL Load Rebalancing Mechanism} first identifies the globally least-loaded worker CPU core with $L_{min}$ tasks and sorts all worker CPU cores in descending order based on their individual queue lengths. To detect load imbalance, the mechanism evaluates each CPU core in the sorted list against the minimum load using two conservative threshold-based conditions: it triggers migration if $L_{min} = 0$ and $L \ge \theta_{idle}$, or if $L > \theta_{load}$ and $L - L_{min} \ge \theta_{gap}$, where $L$ is the queue length of the currently evaluated CPU core. In our implementation, we set $\theta_{idle}=2$, $\theta_{load}=10$, and $\theta_{gap}=5$ to avoid triggering migration for minor queue fluctuations and to rebalance only when idle capacity or substantial localized congestion exists. These thresholds were selected by a parameter sweep on the training workload only. For each qualifying CPU core, the mechanism determines the number of tasks to migrate, defined as half the difference between $L$ and $L_{min}$, and dequeues these tasks from the back of its queue. Finally, the load balancer executes a single batched inference request for all collected tasks through the Core\_Net in the Python-side RL agent, and distributes them to destination worker CPU cores based on the sequential decisions returned by the RL agent.

\textbf{Core\_Net and Slice\_Net} are the two value networks used by the synthesized scheduler. Core\_Net selects destination CPU cores for tasks accumulated in the global queue. This decision has a relatively large discrete action space, since the agent chooses among worker cores. We therefore use a Dueling Deep Q-Network (Dueling DQN)~\cite{wang2016dueling} for Core\_Net, which can improve learning when many actions have similar values. Slice\_Net selects the preemption time slice for worker tasks. Its action space is much smaller, consisting of five candidate time slices. Four of these candidates, 32\,ms, 174\,ms, 390\,ms, and 1,633\,ms, correspond to the 25th, 50th, 75th, and 90th percentiles of our sampled workload durations and are shared by both objectives. The two objectives differ only in the fifth, extreme candidate. For the latency-oriented scheduler we add an aggressive 8\,ms slice, the 10th percentile of workload durations. A short slice forces running tasks to give up the CPU quickly, so a newly arrived task waits less before its first run, which is exactly what the latency objective rewards. For the runtime-oriented scheduler we instead replace this aggressive option with non-preemptive run-to-completion. Here the trade-off runs the opposite way: short slices keep interrupting tasks and switching them between cores, so each task takes longer to finish. Running a task to completion avoids these interruptions and keeps its data on the same core, which in turn lowers the total runtime. Since this is a small discrete action space, we use a simpler Deep Q-Network (DQN)~\cite{mnih2013playing} for Slice\_Net to reduce model complexity.

\textbf{State Space and Normalization.} Both networks observe a shared RL state consisting of the number of started, unstarted, and finished tasks, together with the accumulated latency, accumulated runtime, and accumulated preemption count. Because these quantities feed the network as inputs (distinct from the reward terms defined below), we normalize them to keep training stable: the three task counts are passed through a logarithmic transformation $f(x)=\log(1+x)$ to compress their dynamic range and dampen the effect of bursty arrivals, while the temporal and event-based metrics (accumulated latency, accumulated runtime, and preemption count) are linearly rescaled by fixed denominators. This maps the inputs into a bounded numerical range and prevents vanishing and exploding gradients caused by high-magnitude raw counts.

\vspace{-0.15cm}
\subsection{Optimizing Accumulated Task Latency}
According to Little's Law ($L=\lambda W$)~\cite{little1961proof}, the long-term average number of tasks waiting in a system ($L$) is strictly proportional to the average waiting time ($W$) given a stable arrival rate ($\lambda$). Inspired by this principle, minimizing the accumulation of waiting tasks in the queue intuitively minimizes the expected latency for each task.
Formally, the total accumulated latency can be formulated as the sum of the waiting times for all tasks:
\begin{equation}
    T_{\text {accumulated task latency}} = \sum_i\left(\operatorname{\text{first execution}}_i-\operatorname{arrival}_i\right)
\end{equation}
Let $N_{\text {unstarted }}(t)$ denote the number of tasks that have arrived but have not yet started executing at time $t$.
The optimization objective is equivalent to minimizing the time integral of all waiting tasks over the scheduling period:
\begin{equation}
\min \sum_i\left(\operatorname{\text{first execution}}_i-\operatorname{arrival}_i\right)=\int_0^T N_{\text {unstarted }}(t) d t
\end{equation}

This single count, however, hides where tasks are waiting. If some cores have far longer queues than others, tasks on those cores must wait behind many earlier tasks (head-of-line blocking) even while other cores have nothing to run. To capture this queue imbalance, we define a single \emph{imbalance} penalty $I(t)$ from the per-core queue lengths $Q_k(t)$ (with mean $\bar{Q}$, standard deviation $\sigma_Q$, and maximum $Q_{\max}$), which is reused by both reward functions:
\begin{equation}
I(t) = w_\sigma\,\frac{\sigma_Q(t)}{\max(1,\bar{Q}(t))} + w_b\,\frac{Q_{\max}(t)}{\max(1,\bar{Q}(t))}.
\label{eq:imbalance}
\end{equation}
The first term is the coefficient of variation of the queue lengths and the second is how much the busiest core exceeds the average; $w_\sigma$ and $w_b$ are objective-specific weights fixed within each experiment. We floor the denominator at one task so that $I(t)$ is well-defined and equals zero when the system is idle (all per-core queues empty). Symmetrically, we define a \emph{waiting-work} penalty $W(t)$ that measures how much work is waiting, from the number of unstarted tasks together with the mean and maximum queue lengths:
\begin{equation}
W(t) = w_1\log\!\left(1+N_{\text{unstarted}}\right) + w_2\log\!\left(1+\bar{Q}\right) + w_3\log\!\left(1+Q_{\max}\right).
\label{eq:waiting}
\end{equation}
All counts are log-scaled so that no single term dominates when queues grow to hundreds of tasks during a burst. The system \emph{pressure} $P(t)$ is then simply the sum of the two penalties:
\begin{equation}
P(t) = W(t) + I(t).
\end{equation}
The number of unstarted tasks leads with the largest weight $w_1$ because, by Little's Law, it is the objective; the queue-length and imbalance terms are secondary signals that discourage the imbalanced states causing head-of-line blocking and starvation. Across all latency-oriented experiments we fix $(w_1, w_2, w_3) = (2.0,\,0.5,\,1.0)$ and $(w_\sigma, w_b) = (0.5,\,0.3)$. These weights were selected on the training workload and are kept unchanged for all held-out workloads.

Finally, we convert this pressure into a per-step difference-based reward. Let $\Phi(t)=-\log(1+P(t))$, where lower pressure corresponds to a larger potential value. We define the reward as the change in this potential between consecutive scheduling steps:
\begin{equation}
r_t = \Phi(t+1) - \Phi(t).
\end{equation}
Each step therefore rewards the agent for reducing pressure. Under the discounted return optimized by DQN, this difference-based reward favors policies that reduce pressure earlier and keep it low throughout the episode.

\vspace{-0.15cm}
\subsection{Optimizing Accumulated Task Runtime}
The runtime of a task is defined as the duration from its first execution on CPU to its final completion. The accumulated runtime across all tasks is formulated as:
\begin{equation}
    T_{\text {accumulated task runtime}} = \sum_i\left(\operatorname{\text{completion}}_i-\operatorname{\text{first execution}}_i\right)
\end{equation}
In our setting, the task runtime can be prolonged by scheduling-induced overheads, such as delays caused by preemption and requeuing. A highly imbalanced queue distribution can cause idle cores to steal tasks from busy cores, leading to frequent preemptions and prolonged runtime. Since the final task runtime is only observed at completion, we use runtime-related signals observable during scheduling: the preemption rate and queue imbalance.

The dominant cause is preemption, so here the primary term is the preemption rate: the number of preemptions $\Delta P(t)$ per scheduling step normalized by the core count $K$:
\begin{equation}
\rho_P(t) = \frac{\Delta P(t)}{K}.
\end{equation}
We normalize by the number of cores rather than by the number of queued tasks so that a step full of preemptions stays clearly distinguishable from a preemption-free step even when the queues are deep. Reusing the same imbalance penalty $I(t)$ from Eq.~\eqref{eq:imbalance} (since a lopsided queue lets idle cores steal tasks from busy ones and triggers extra preemptions), the runtime pressure has the same \emph{primary + imbalance} form as before:
\begin{equation}
P_{\text{exec}}(t) = w_p\,\rho_P(t) + I(t).
\end{equation}
The preemption rate carries by far the largest weight because it is the most direct driver of runtime; the imbalance penalty plays the same secondary, locality-preserving role as in the latency reward, with smaller weights. Across all runtime-oriented experiments we fix $w_p = 4.0$ and $(w_\sigma, w_b) = (0.2,\,0.05)$. These weights were selected on the training workload and are kept unchanged for all held-out workloads.

We map this pressure to a bounded per-step reward with a logarithm:
\begin{equation}
r_t = -\log\!\left(1 + P_{\text{exec}}(t)\right),
\end{equation}
which keeps rewards in a stable range and yields diminishing penalties as pressure grows. Maximizing $\sum_t r_t$ trains a policy that avoids needless preemption and severe queue imbalance, preserving locality and shortening accumulated task runtime.

\vspace{-0.15cm}
\subsection{Training Convergent Models}\label{subsec:training}
\begin{figure}[h]
    \vspace{-0.3cm}
    \centering    
    \includegraphics[width=0.9\linewidth]{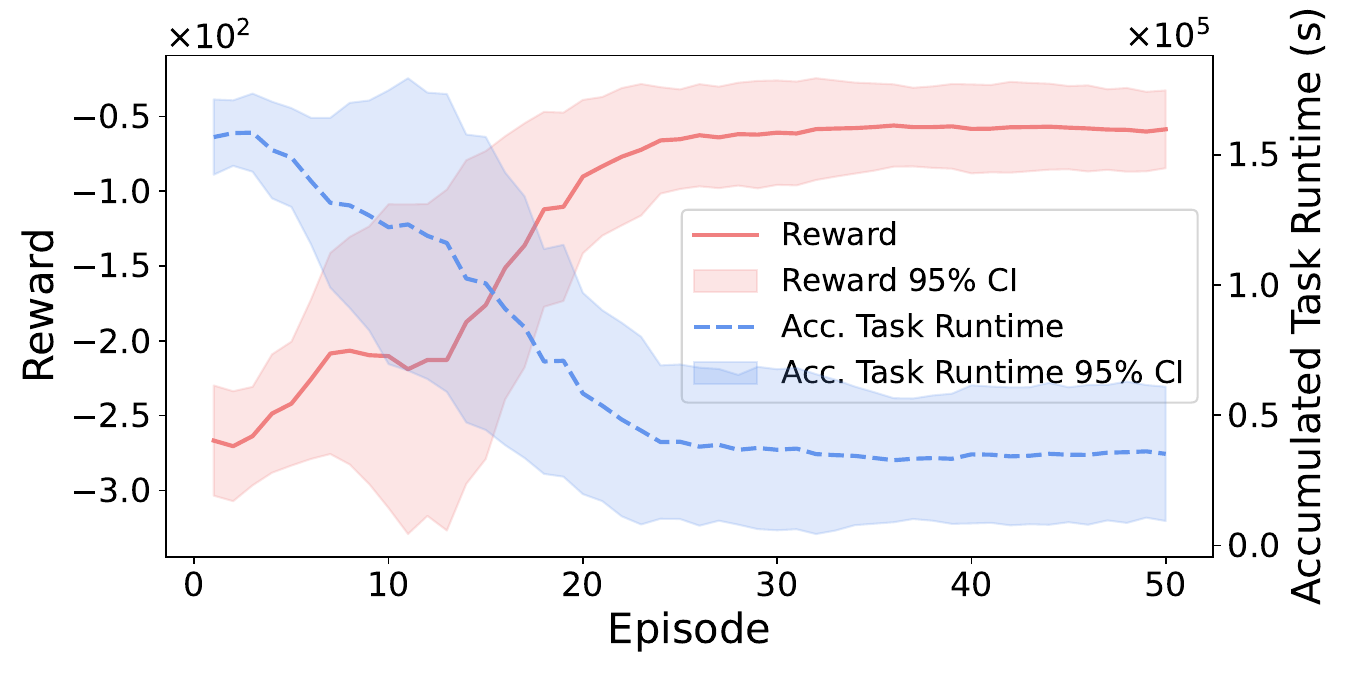}
    \vspace*{-0.4cm}
    \caption{Reward and accumulated runtime over 50 training episodes, with 95\% confidence bands across five random seeds. Both stabilize within 50 episodes, indicating convergence.}
    \vspace*{-0.5cm}
    \label{fig:training_exec}
\end{figure}

We train each scheduler on the training workload for 50 episodes using five independent random seeds. \Cref{fig:training_exec} shows the reward and the corresponding runtime metrics with 95\% confidence intervals over the 50 training episodes. The results show that both reward and latency/runtime stabilize within 50 episodes, indicating that the synthesized scheduler can learn convergent policies for core assignment, preemption, and load rebalancing. Training is a one-time cost: each episode takes 6--8 minutes, so training one policy for 50 episodes takes 5--7 hours on the setup in~\Cref{subsec:setup}.

For each episode, we compute the mean and confidence interval across the five runs with the same episode index. The standard error of the mean (SEM) is computed as $\text{SEM} = \frac{\sigma}{\sqrt{n}}$, where $\sigma$ is the sample standard deviation across runs and $n=5$ is the number of runs. We use a normal-approximation 95\% confidence interval, computed as $\text{mean} \pm 1.96 \times \text{SEM}$. For visualization, we apply a 5-episode moving average to smooth both mean curves and confidence bands.

\vspace{-0.15cm}
\subsection{Exploring Latency--Runtime Pareto Space}\label{subsec:pareto}
Training one model per operating point is impractical. Instead, we synthesize two policies, one latency-oriented and one runtime-oriented, which share the same architecture and placement action space and differ only in their reward, so that their operating points are directly comparable. We then show that a single trained policy can be tuned at deployment time to trace out the latency--runtime trade-off, without retraining. We take the runtime-optimized Dueling DQN policy and, at evaluation time, hold its placement behavior fixed while tuning two parameters that control preemption. The first is the preemption time slice: a short slice interrupts running tasks quickly, so newly arrived tasks reach the CPU sooner (lower accumulated latency) at the price of more preemptions and context switches (higher accumulated runtime), whereas a long slice lets tasks run largely to completion (lower runtime) at the price of longer waiting (higher latency). The slice alone, however, moves the operating point in discrete jumps: because the workload's execution times are clustered, shrinking the slice pushes whole groups of tasks past the preemption threshold at once, leaving gaps in the trade-off. To fill these gaps, we tune a second parameter: a global budget on the total number of task preemptions per episode. Varying this budget changes the total preemption count, and hence accumulated latency and runtime. Tuning both parameters yields a near-continuous set of operating points spanning the latency--runtime trade-off, as seen in Figure~\ref{fig:pareto_scatter_middle_dense_seeds}.

A natural question is whether a heuristic scheduler such as CFS can reach this near-continuous set of operating points simply by tuning the preemption parameters. In our exploration, we hold the learned placement policy fixed and vary only preemption. Therefore, the set of operating points is decided not only by the preemption parameters but also by the placement policy. CFS places and migrates tasks with a fairness-driven, serverless-agnostic load balancer, which prolongs runtime through preemptions. Thus, tuning its time slice still traces a set of operating points in the high-runtime region. Furthermore, CFS also exposes no budget on the total number of preemptions. Its preemptions follow fairness among tasks instead of a tunable global budget, leaving it unable to fill the gaps left by the slice alone. Thus, tuning CFS yields a set of operating points in the high-runtime region instead of the near-continuous set our synthesized schedulers can reach.

This limitation is not specific to CFS. Each heuristic scheduler that we evaluate is designed around a single objective and therefore occupies one region of the trade-off: ALPS~\cite{fu2024alps}, CFS, and EEVDF preempt aggressively and occupy the high-runtime region; FIFO and SFS~\cite{fu2022sfs} preempt much less and occupy the high-latency region; Hybrid~\cite{zhao2024serverless} occupies an intermediate region. These schedulers expose tunable parameters, but none of them decouples placement from preemption the way synthesized schedulers do. Our synthesized schedulers decouple placement from preemption: they keep a learned placement policy fixed while independently tuning the preemption slice and a global preemption budget. This separation allows deployment-time tuning to expose a near-continuous set of operating points without retraining the scheduler. At deployment, the operator selects an operating point by setting the two preemption parameters to meet its SLO and cost targets. A provider can also run pools of nodes at different operating points and price them accordingly, letting clients pick the pool that matches their needs.

\begin{figure}[t]
    \centering
    \includegraphics[width=0.65\linewidth]{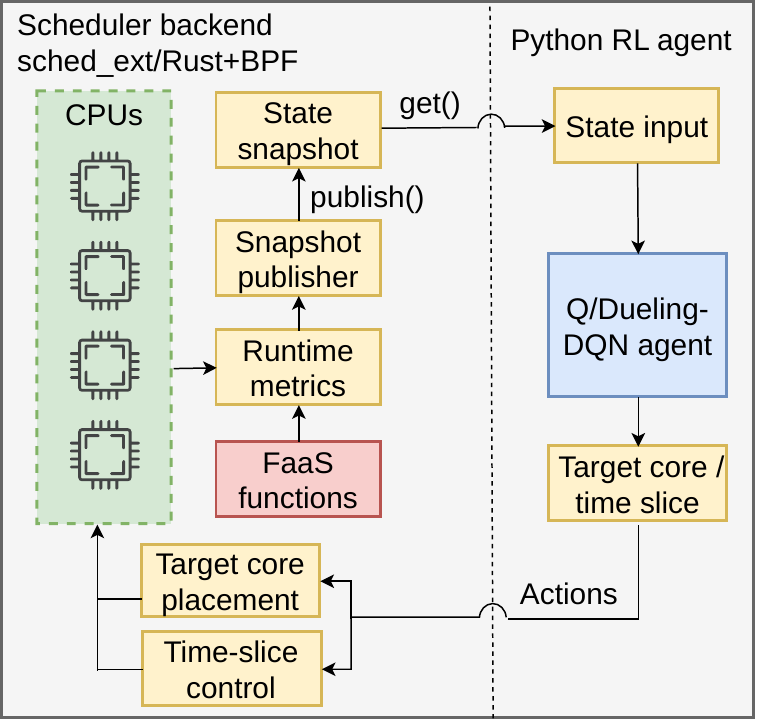}
    \vspace*{-0.4cm}
    \caption{Synthesized scheduler communication pattern: the \texttt{sched\_ext} Rust+BPF backend (left) publishes a state snapshot of runtime metrics to the Python RL agent (right); the Q/Dueling-DQN agent returns target-core and time-slice decisions, which the backend applies as core placement and time-slice control.}
    \label{fig:architecture}
    \vspace*{-0.5cm}
\end{figure}

\vspace{-0.15cm}
\subsection{Implementation}\label{subsec:implementation}
We implement synthesized schedulers on top of  \texttt{sched\_ext}~\cite{schedext}, a modern framework for implementing custom Linux schedulers with BPF. Our implementation follows a high-level architecture: a scheduler backend, a Python-based RL agent, and a synchronization bridge between the scheduler and the learning logic. Figure~\ref{fig:architecture} shows the interaction of these components.

\textbf{\texttt{sched\_ext} backend.} Synthesized schedulers use a Rust user-space component together with a BPF scheduling component. The BPF component is the kernel-side interface to \texttt{sched\_ext}: it receives scheduling callbacks, forwards runnable-task events to the Rust scheduler, and applies scheduling decisions by dispatching tasks into \texttt{sched\_ext} dispatch queues (DSQs). A DSQ is the kernel-side dispatch queue where CPU cores pick runnable tasks from. The Rust user-space maintains a global queue and per-core queues, invokes the RL agent to make placement and rebalancing decisions, and returns target cores and time slices to the kernel  interface.

\textbf{RL agent.} The RL agent is implemented in Python using PyTorch~\cite{paszke2019pytorch}, and produces both CPU-placement decisions and time-slice decisions using Deep Q-learning and Dueling Deep Q-Networks. Python communicates with the Rust scheduler using \textit{PyO3}~\cite{pyo3}. The scheduler exposes state snapshots to Python, including queue lengths, numbers of started, unstarted, and finished tasks, latency and runtime statistics, and preemption information. The agent uses these snapshots to compute rewards and update the Q-network.

\textbf{Synchronization.} To keep Python inference off the scheduler fast path, the RL agent runs on a dedicated core outside the worker set, so its  inference and training never compete with scheduled tasks. Each round, the Rust scheduler publishes the latest state snapshot into a shared in-process slot that the embedded Python interpreter reads, and then invokes the agent through \textit{PyO3} to obtain placement and time-slice actions. Decisions run at two frequencies to limit how often Python is called. Placement is updated every 20\,ms, while the time slice and load rebalancing are updated every 500\,ms.

\vspace{-0.15cm}
\section{Experimental Setup}
In this section, we first describe the hardware and software setup (\Cref{subsec:setup}). We then introduce the workloads and metrics used in our experiments (\Cref{subsec:workload}).

\vspace{-0.15cm}
\subsection{Hardware and Software Setup}\label{subsec:setup}
We use a server provided by CloudLab~\cite{duplyakin2019design} with two Intel Xeon Gold 6142 CPUs (2.60~GHz, 16 cores each with hyper-threading) providing 64 CPU cores in total, and 384~GB of DRAM. We dedicate 50 of these cores to the \texttt{sched\_ext} worker pool; this leaves headroom for the user-space scheduler, the RL agent, and background system activity, so that the agent's PyTorch inference and training do not compete with scheduled tasks. All main experiments run the scheduler policy exclusively on CPUs, with PyTorch CUDA support disabled. GPUs are evaluated only in a separate microbenchmark comparing policy-inference latency with CUDA enabled versus disabled on an NVIDIA A30. \textbf{OS:} Ubuntu 26.04 LTS (Linux 7.0.0-10-generic) with \texttt{sched\_ext} enabled. For a fair comparison, we reimplemented all evaluated schedulers in \texttt{sched\_ext}, following their original designs.

\vspace{-0.15cm}
\subsection{Workloads and Metrics}\label{subsec:workload}
We train on serverless function arrival patterns extracted from the real-world Microsoft Azure FaaS trace~\cite{shahrad2020serverless}, following the trace-extraction methodology of prior work~\cite{fu2022sfs, fu2024alps, zhao2024serverless}. The training workload is the arrival sequence from the first two minutes of Day~1, whose execution durations are highly heterogeneous (8--25{,}280\,ms), preserving the high variance of production serverless environments. To test generalization, we evaluate the trained policy zero-shot on three held-out workloads (Workload~1--3): a high-traffic noon period of Day~1 (intra-day shift), the first two minutes of Day~2 (inter-day shift), and a four-minute period of Day~10 (both), in Table~\ref{tab:workloads}.

\begin{table}[t]
    \vspace{-0.3cm}
    \caption{Summary of training and test workloads.}
    \vspace{-0.4cm}
    \centering
    \footnotesize
    \setlength{\tabcolsep}{4pt}
    \begin{tabular*}{0.95\columnwidth}{@{\extracolsep{\fill}}l l r p{2.5cm}@{}}
        \toprule
        \textbf{Workload} & \textbf{Period} & \textbf{Invocations} & \textbf{Generalization} \\
        \midrule
        Training & Day01, midnight & 8,285 & -- \\
        Workload~1 & Day02, midnight & 8,627 & day-to-day \\
        Workload~2 & Day01, noon & 9,463 & time-of-day \\
        Workload~3 & Day10, evening & 18,155 & time-of-day \& day-to-day \\
        Workload~4 & Day02, midnight & 8,627 & heterogeneous \\
        Workload~5 & Day01, noon & 9,463 & heterogeneous \\
        Workload~6 & Day10, evening & 18,155 & heterogeneous \\
        \bottomrule
    \end{tabular*}
    \label{tab:workloads}
     \vspace*{-0.5cm}
\end{table}

To check that our results are not tied to a single function implementation, we build three heterogeneous workloads (Workload~4--6) spanning the resource dimensions from prior characterization studies (e.g., SeBS~\cite{copik2021sebs}): CPU-bound (recursive Fibonacci, our calibrated baseline), memory-bandwidth-bound (sequential array scan/copy/write), and I/O-blocking (a short nanosleep with cached file read and checksum). The three stress different scheduling paths: CPU-bound tasks drive time-slice and priority accounting; memory-bound tasks stay runnable but have contention-inflated service times; and I/O-blocking tasks are the only ones that transition between runnable and blocked states. We mix them 70\%/20\%/10\%, so most load exercises the CPU-bound behavior our scheduler targets while the memory and I/O fractions stress-test service-time uncertainty and blocked-task transitions. Our design has no dedicated I/O mechanism, but blocked tasks still benefit indirectly: on wakeup, load balancing can place them on a less busy core to resume sooner.

Because serverless platforms continuously execute function invocations, we characterize scheduling performance by two metrics accumulated over a workload. For each invocation, its \emph{latency} is the time from arrival to when it first runs, and its \emph{runtime} is the time from its first run to completion. Runtime is wall-clock time rather than CPU time: it includes the time a task spends preempted and waiting to resume. We report \emph{accumulated task latency} and \emph{accumulated task runtime} as the sums over all invocations in the workload, for both of which lower is better. Accumulated latency is critical for deadline-constrained, user-facing workloads, whereas accumulated runtime determines billed cost under pay-as-you-go pricing. We report cost as accumulated runtime normalized to CFS.

\begin{figure}[tp]
    \vspace{-0.35cm}
    \centering    
    \includegraphics[width=0.85\linewidth]{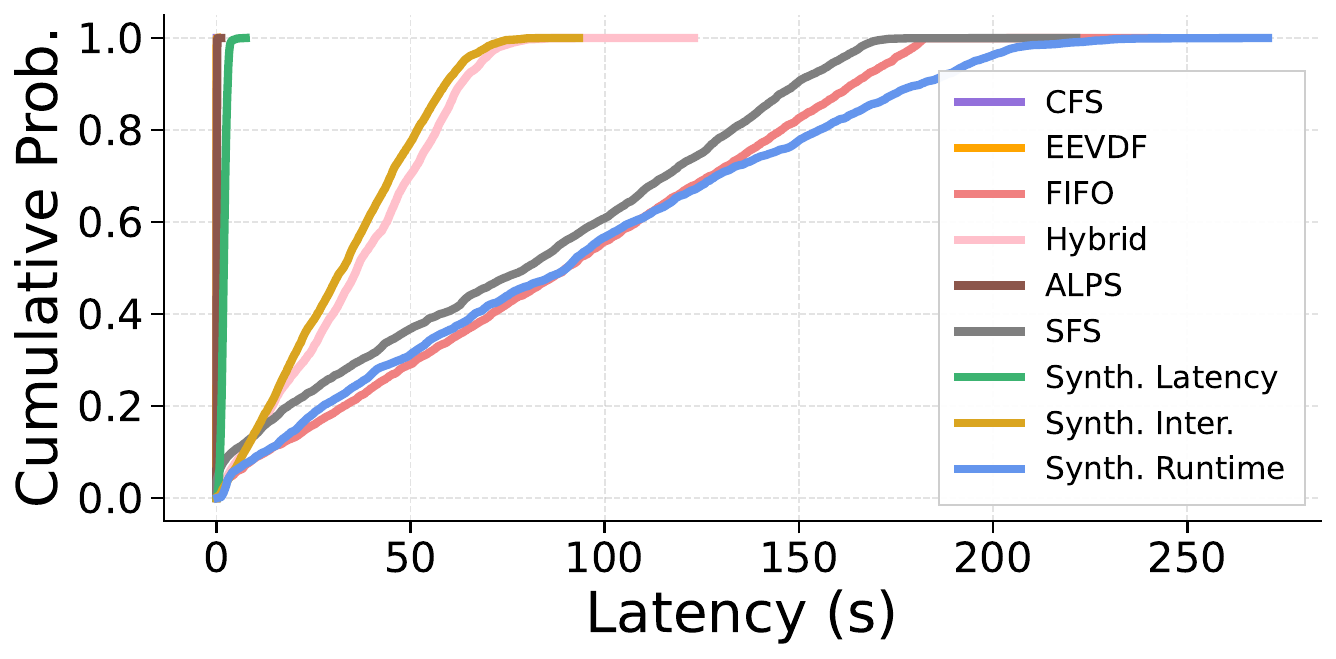}
    \vspace*{-0.45cm}
    \caption{CDFs of response latency for a workload extracted from the Microsoft Azure trace under CFS, EEVDF, FIFO, Hybrid~\cite{zhao2024serverless}, ALPS~\cite{fu2024alps}, SFS~\cite{fu2022sfs}, and three synthesized scheduling policies on 50 CPU cores. Synthesized Latency closely tracks the low-latency schedulers (CFS, EEVDF, ALPS), showing it stays in its target region on an unseen workload.}
    \vspace*{-0.2cm}
    \label{fig:response_test_day02}
\end{figure}

\begin{figure}[tp]
    \centering
    \includegraphics[width=0.85\linewidth]{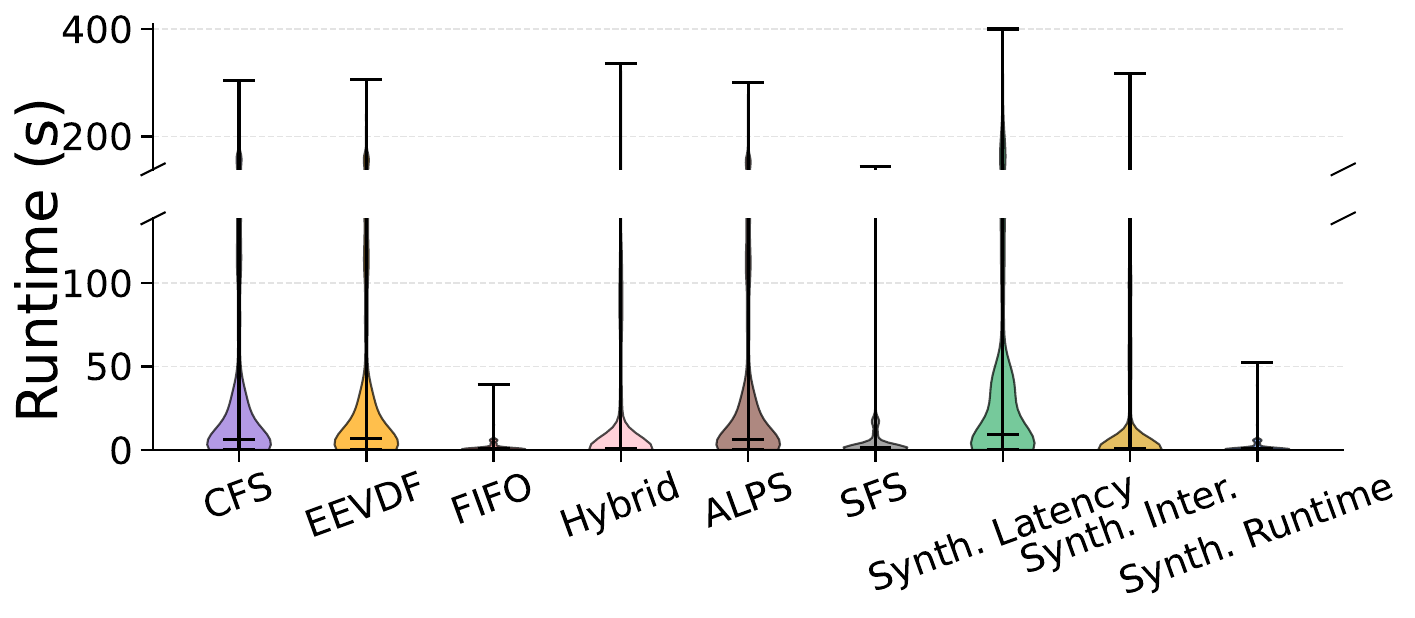}
    \vspace*{-0.45cm}
    \caption{Task runtime distributions for a workload extracted from the Microsoft Azure trace under CFS, EEVDF, FIFO, Hybrid~\cite{zhao2024serverless}, ALPS~\cite{fu2024alps}, SFS~\cite{fu2022sfs}, and three synthesized scheduling policies on 50 CPU cores. Synthesized Runtime overlaps FIFO, the practical runtime lower bound.}
    \vspace*{-0.5cm}
    \label{fig:execution_test_day02}
\end{figure}

\vspace{-0.15cm}
\section{Evaluation}

We evaluate the schedulers synthesized by our system along three axes: effectiveness, generality, and the runtime cost of the synthesis architecture. We first establish that the learned policies hold up beyond the conditions they were trained on: \Cref{subsec:generalization} shows they remain effective when the workload distribution shifts across intra-day and inter-day traffic patterns, and \Cref{subsec:heterogeneous} shows the frozen models transfer to heterogeneous workloads without retraining. We then characterize the resulting behavior and its price: \Cref{subsec:throughput} analyzes task-completion throughput, and \Cref{subsec:overhead} quantifies the per-task scheduling overhead, confirming the approach is practical on the decision path. Finally, \Cref{subsec:pareto_results} turns to our central claim: by tuning the preemption parameters at deployment time, our schedulers with learned task placement policy span the latency--runtime trade-off, and we use this view to examine the provider--client cost and the fairness--latency trade-offs across schedulers.

\begin{table}[t]
  
    \caption{Performance comparison on heterogeneous workloads (Workload~4--6). Latency and runtime are accumulated values across all tasks. Cost is normalized to CFS within each workload. Lower is better for all metrics. Each synthesized policy holds its target operating region. Values are mean $\pm$ standard deviation over 5 independent runs.}
    \vspace{-0.4cm}
    \centering
    \footnotesize
    \setlength{\tabcolsep}{4pt}
    \begin{tabular*}{0.99\columnwidth}{@{\hspace{\tabcolsep}\extracolsep{\fill}}l r r r}
        \toprule
        \textbf{Policy} & \textbf{Latency (s)}
        & \textbf{Runtime (s)} & \textbf{Cost} \\
        \midrule
        \textit{Workload~4} & & & \\
        \rowcolor{latencyregion} \textbf{Low-latency region} & & & \\
        \rowcolor{latencyregion} CFS & 173 $\pm$ 2 & 209,279 $\pm$ 2,247 & 1.0000 \\
        \rowcolor{latencyregion} EEVDF & 168 $\pm$ 0.4 & 212,022 $\pm$ 889 & 1.0132 $\pm$ 0.0104 \\
        \rowcolor{latencyregion} ALPS & 567 $\pm$ 6 & 204,328 $\pm$ 1,594 & 0.9765 $\pm$ 0.0156 \\
        \rowcolor{latencyregion} Synthesized Latency & 14,524 $\pm$ 668 & 308,601 $\pm$ 11,137 & 1.4748 $\pm$ 0.0592 \\
        \rowcolor{intermediateregion} \textbf{Intermediate region} & & & \\
        \rowcolor{intermediateregion} Hybrid & 219,258 $\pm$ 1,151 & 152,015 $\pm$ 881 & 0.7264 $\pm$ 0.0069 \\
        \rowcolor{intermediateregion} Synthesized Inter. & 194,694 $\pm$ 3,485 & 143,681 $\pm$ 750 & 0.6866 $\pm$ 0.0081 \\
        \rowcolor{runtimeregion} \textbf{Low-runtime region} & & & \\
        \rowcolor{runtimeregion} FIFO & 631,506 $\pm$ 5,694 & 75,814 $\pm$ 601 & 0.3623 $\pm$ 0.0056 \\
        \rowcolor{runtimeregion} SFS & 541,086 $\pm$ 19,282 & 84,253 $\pm$ 1,862 & 0.4027 $\pm$ 0.0123 \\
        \rowcolor{runtimeregion} Synthesized Runtime & 657,675 $\pm$ 7,182 & 69,184 $\pm$ 914 & 0.3306 $\pm$ 0.0059 \\
        \addlinespace[4pt]
        \textit{Workload~5} & & & \\
        \rowcolor{latencyregion} \textbf{Low-latency region} & & & \\
        \rowcolor{latencyregion} CFS & 252 $\pm$ 1 & 323,220 $\pm$ 1,110 & 1.0000 \\
        \rowcolor{latencyregion} EEVDF & 247 $\pm$ 2 & 324,492 $\pm$ 2,471 & 1.0039 $\pm$ 0.0070 \\
        \rowcolor{latencyregion} ALPS & 833 $\pm$ 5 & 313,711 $\pm$ 1,018 & 0.9706 $\pm$ 0.0049 \\
        \rowcolor{latencyregion} Synthesized Latency & 16,288 $\pm$ 646 & 402,266 $\pm$ 6,471 & 1.2445 $\pm$ 0.0168 \\
        \rowcolor{intermediateregion} \textbf{Intermediate region} & & & \\
        \rowcolor{intermediateregion} Hybrid & 325,889 $\pm$ 1,980 & 242,447 $\pm$ 1,250 & 0.7501 $\pm$ 0.0061 \\
        \rowcolor{intermediateregion} Synthesized Inter. & 284,653 $\pm$ 2,091 & 225,268 $\pm$ 881 & 0.6970 $\pm$ 0.0042 \\
        \rowcolor{runtimeregion} \textbf{Low-runtime region} & & & \\
        \rowcolor{runtimeregion} FIFO & 985,991 $\pm$ 4,176 & 110,750 $\pm$ 587 & 0.3426 $\pm$ 0.0011 \\
        \rowcolor{runtimeregion} SFS & 870,321 $\pm$ 12,469 & 118,430 $\pm$ 1,116 & 0.3664 $\pm$ 0.0034 \\
        \rowcolor{runtimeregion} Synthesized Runtime & 1,022,983 $\pm$ 18,546 & 100,151 $\pm$ 2,695 & 0.3098 $\pm$ 0.0080 \\
        \addlinespace[4pt]
        \textit{Workload~6} & & & \\
        \rowcolor{latencyregion} \textbf{Low-latency region} & & & \\
        \rowcolor{latencyregion} CFS & 733 $\pm$ 12 & 764,200 $\pm$ 3,537 & 1.0000 \\
        \rowcolor{latencyregion} EEVDF & 732 $\pm$ 4 & 766,990 $\pm$ 2,212 & 1.0037 $\pm$ 0.0070 \\
        \rowcolor{latencyregion} ALPS & 2,391 $\pm$ 6 & 732,933 $\pm$ 2,330 & 0.9591 $\pm$ 0.0038 \\
        \rowcolor{latencyregion} Synthesized Latency & 41,778 $\pm$ 2,076 & 940,402 $\pm$ 19,288 & 1.2305 $\pm$ 0.0226 \\
        \rowcolor{intermediateregion} \textbf{Intermediate region} & & & \\
        \rowcolor{intermediateregion} Hybrid & 925,879 $\pm$ 3,522 & 522,586 $\pm$ 4,468 & 0.6838 $\pm$ 0.0061 \\
        \rowcolor{intermediateregion} Synthesized Inter. & 762,273 $\pm$ 44,265 & 496,935 $\pm$ 8,440 & 0.6503 $\pm$ 0.0125 \\
        \rowcolor{runtimeregion} \textbf{Low-runtime region} & & & \\
        \rowcolor{runtimeregion} FIFO & 2,556,550 $\pm$ 34,911 & 271,454 $\pm$ 2,397 & 0.3552 $\pm$ 0.0023 \\
        \rowcolor{runtimeregion} SFS & 2,294,877 $\pm$ 46,732 & 279,546 $\pm$ 9,007 & 0.3658 $\pm$ 0.0131 \\
        \rowcolor{runtimeregion} Synthesized Runtime & 2,559,328 $\pm$ 26,399 & 241,361 $\pm$ 5,444 & 0.3158 $\pm$ 0.0077 \\
        \bottomrule
    \end{tabular*}
    \vspace*{-0.35cm}
    \label{tab:workload456}
\end{table}

\vspace{-0.15cm}
\subsection{Generalization to Unseen Arrival Pattern}\label{subsec:generalization}

We first evaluate the robustness of our learned schedulers under workload shifts and previously unseen arrival patterns. Because each scheduler is synthesized from a fixed set of training workloads, a key question is whether the learned policy remains effective once the workload distribution changes at deployment time.

In RL evaluation, the trained policy is executed without exploration (i.e., $\epsilon = 0$ in $\epsilon$-greedy), and performance metrics such as total accumulated latency and total runtime are collected. We evaluate our synthesized scheduler on unseen test workloads sampled from different time windows of the Azure trace. These workloads represent various real-world traffic patterns, including intra-day shifts (e.g., noon vs. midnight) and inter-day variations (e.g., Day~1 vs. Day~2); see~\Cref{subsec:workload} for details.

Figures~\ref{fig:response_test_day02} and \ref{fig:execution_test_day02} report results for Workload~1, one of the held-out test workloads defined in \Cref{subsec:workload}, using policies trained on the training workload from the same Azure trace. For this workload, we plot the CDF of per-task response latency and the distribution of per-task runtime across all schedulers. Despite the workload shift, each synthesized policy stays within its intended operating region. In the response-latency CDF, Synthesized Latency clusters with the low-latency schedulers (CFS, EEVDF, ALPS), completing nearly all tasks within a few seconds and staying far ahead of the higher-latency schedulers (Hybrid, FIFO, SFS). In the runtime distribution, Synthesized Runtime overlaps FIFO, the practical runtime lower bound. Between these two extremes, Synthesized Intermediate (selected in \Cref{subsec:pareto}) balances the two objectives, trading a moderate amount of latency for substantially lower runtime.

\vspace{-0.15cm}
\subsection{Heterogeneous Workload}\label{subsec:heterogeneous}

\begin{table}[t]
    \caption{Function completion timing across scheduling policies (0--120s / 0--240s: share completed within each window; Tail\,$>$240s: share finishing after 240s; $T_{99\%}$: time to complete 99\%). Synthesized policies stay comparable to the baselines on completion throughput.}
    \vspace{-0.4cm}
    \centering
    \footnotesize
    \setlength{\tabcolsep}{4pt}
    \begin{tabular*}{0.98\columnwidth}{@{\extracolsep{\fill}}l r r r r@{}}
        \toprule
        \textbf{Policy} & \textbf{0--120s} & \textbf{0--240s} & \textbf{Tail $>$240s} & \textbf{$T_{99\%}$} \\
        \midrule
        CFS & 69.3\% & 89.2\% & 10.8\% & 359.2s \\
        EEVDF & 69.2\% & 89.4\% & 10.6\% & 361.5s \\
        FIFO & 33.3\% & 66.0\% & 34.0\% & 358.5s \\
        Hybrid & 48.6\% & 88.7\% & 11.3\% & 422.1s \\
        ALPS & \textbf{69.7\%} & 89.8\% & 10.2\% & 352.1s \\
        SFS & 36.5\% & 67.8\% & 32.2\% & \textbf{351.9s} \\
        Synthesized Latency & 66.4\% & 85.9\% & 14.1\% & 413.2s \\
        Synthesized Intermediate & 50.9\% & \textbf{90.2\%} & \textbf{9.8\%} & 361.9s \\
        Synthesized Runtime & 32.8\% & 66.4\% & 33.6\% & 377.0s \\
        \bottomrule
    \end{tabular*}
    \vspace{-0.5cm}
    \label{tab:throughput}
\end{table}

To test whether the learned policies transfer beyond a single function type, we evaluate the frozen models, without retraining, on three heterogeneous workloads (Workload~4--6) that mix CPU-bound, memory-bandwidth-bound, and I/O-blocking functions (\Cref{subsec:workload}). All three synthesized policies retain their target operating regions on these mixed workloads, confirming that the learned behavior is not specific to the original CPU-bound Fibonacci implementation. Table~\ref{tab:workload456} quantifies this behavior. The lowest accumulated task latency comes from the low-latency schedulers CFS, EEVDF, and ALPS, which Synthesized Latency does not match. Against every other scheduler (Hybrid, FIFO, SFS), however, Synthesized Latency reduces accumulated latency by more than an order of magnitude on all three workloads. It thus stays in the low-latency region, at the cost of higher runtime than CFS, EEVDF, and ALPS. Synthesized Intermediate dominates the Hybrid heuristic it targets on all three workloads, cutting accumulated task latency by 11\% to 18\% and accumulated task runtime and cost by 5\% to 7\%. Synthesized Runtime reaches the low-runtime region, matching FIFO on both accumulated task runtime and cost.

Although non-preemptive FIFO is expected to minimize accumulated task runtime, this expectation holds only for tasks that run to completion without blocking after their first dispatch. Decomposing accumulated task runtime by function type confirms this distinction: for continuously runnable CPU- and memory-bandwidth-bound tasks, FIFO achieves 2--11\% lower accumulated task runtime than Synthesized Runtime. I/O-blocking tasks, in contrast, repeatedly block on read operations and must be re-dispatched after each wakeup. Under FIFO, a woken task is requeued at the tail and, when the system is oversubscribed, waits behind other work before it can resume. On the I/O-blocking tasks, this waiting makes FIFO's first-run-to-completion times 12--15\% longer than Synthesized Runtime's, which places and rebalances woken tasks on less-loaded cores. Since I/O-blocking tasks account for roughly 79\% of accumulated task runtime, this reduction outweighs FIFO's small advantage on the non-blocking tasks, allowing Synthesized Runtime to achieve lower total accumulated runtime and cost.

\vspace{-0.15cm}
\subsection{Throughput}\label{subsec:throughput}

Table~\ref{tab:throughput} summarizes the completion-time distribution across scheduling policies. CFS, EEVDF, and ALPS complete around 69\% of tasks within 120 seconds and have short tails, with only about 10\% completing after 240 seconds. In contrast, FIFO completes only 33\% within 120 seconds, with 34\% completing after 240 seconds. The latency-optimized synthesized scheduler closely follows the low-latency schedulers over most of the distribution, completing 66\% of tasks within 120 seconds and 86\% within 240 seconds, substantially ahead of the FIFO-style schedulers (FIFO, SFS, and Synthesized Runtime), which complete only about one third within 120 seconds. Its tail is slightly longer than the best low-latency baseline: 14\% of tasks complete after 240 seconds, and $T_{99\%}$ is roughly 413 seconds versus about 352 seconds for ALPS. The intermediate synthesized scheduler, which targets the same operating region as the Hybrid heuristic, has a tail (10\%) and $T_{99\%}$ (362 seconds) comparable to CFS, EEVDF, and ALPS. The runtime-optimized synthesized scheduler exhibits a completion distribution similar to FIFO: since it optimizes accumulated task runtime, the learned policy uses longer preemption time slices and performs fewer preemptions.

\begin{table}[tp]
    % \vspace{-0.3cm}
    \caption{Per-task scheduling overhead of the synthesized schedulers, split into scheduler--agent communication and RL inference. Overhead of $\mu$s scale is sufficiently small for serverless scheduling decisions.}
    \vspace{-0.4cm}
    \centering
    \footnotesize
    \setlength{\tabcolsep}{3pt}
    \begin{tabular*}{0.95\columnwidth}{@{\extracolsep{\fill}}l r r r@{}}
        \toprule
         & \textbf{Comm} & \textbf{RL inference} & \textbf{Total} \\
        \midrule
        Scheduling overhead per task ($\mu$s)             & 0.67  & 134.50 & 135.17 \\
        Share of whole workload (\%)      & 0.002 & 0.40   & 0.40 \\
        \bottomrule
    \end{tabular*}
    \label{tab:overhead_comparison}
    \vspace{-0.5cm}
\end{table}

\vspace{-0.15cm}
\subsection{Overhead Analysis}\label{subsec:overhead}
Table~\ref{tab:overhead_comparison} reports the per-task scheduling overhead of
the synthesized schedulers on the \texttt{sched\_ext} backend, decomposed into
scheduler--agent communication (Comm) and RL inference. The total per-task
overhead is about 135\,$\mu$s and is dominated by RL inference
(134.50\,$\mu$s/task), while communication between the scheduler and the RL agent
adds only 0.67\,$\mu$s/task. Aggregated over an entire episode, this amounts to
just 0.40\% of the whole workload's runtime (0.002\% for communication), as the
agent runs on a single dedicated core alongside the 50 worker cores. The per-task
cost is thus small enough to support online scheduling decisions for serverless
functions, whose runtimes are typically on the order of milliseconds to seconds.

We further evaluate whether GPU execution reduces the online inference overhead of the synthesized scheduler. This experiment only changes the PyTorch execution device while keeping the same trained weights, model architecture, batch size, and torch build. We run a controlled per-decision microbenchmark on an idle agent with batch size one, matching the inference pattern used during evaluation. The GPU setting uses an NVIDIA A30, while the CPU setting disables CUDA. Table~\ref{tab:overhead_ratio} shows that GPU execution increases inference overhead for both decision paths. CPU-core assignment is 2.57$\times$ slower on the GPU, and preemption time-slice updates are 2.98$\times$ slower. The per-task scheduling overhead follows the CPU-core assignment path because each task requires one placement inference, while the global time-slice update is amortized across tasks. Communication remains unchanged because transferring the scheduler state to the RL agent and returning the selected action do not depend on the inference device. These results indicate that GPU execution is not beneficial for our online scheduler: the policy networks are small and each decision uses a single state vector, so kernel-launch, host--device transfer, and synchronization overhead dominate the computation. Therefore, no extra costly hardware is needed for performing inference during serverless scheduling.

\vspace{-0.15cm}
\subsection{Latency--Cost Trade-off in Practice}\label{subsec:pareto_results}

In our serverless workload setting, the Pareto space is defined by two objectives: accumulated task latency and accumulated task runtime. Accumulated latency is critical for deadline-constrained workloads, whereas accumulated runtime directly determines user cost under the pay-as-you-go pricing model. Reducing latency often requires aggressive scheduling decisions, such as earlier preemption, which can increase total runtime, as observed in CFS, EEVDF, and ALPS. Conversely, minimizing accumulated runtime can reduce user-facing cost but may adversely affect deadline-constrained tasks, as demonstrated by FIFO and SFS. Therefore, single-objective evaluation cannot fully characterize scheduler behavior.

\begin{table}[t]
    % \vspace{-0.3cm}
    \caption{GPU/CPU overhead ratio for scheduler decision paths on an NVIDIA A30. Values larger than one indicate that GPU execution is slower, so no accelerator is needed for online inference.}
    \vspace{-0.4cm}
    \centering
    \footnotesize
    \setlength{\tabcolsep}{3pt}
    \begin{tabular*}{0.95\columnwidth}{@{\extracolsep{\fill}}l r r r@{}}
        \toprule
        \textbf{Metric} & \textbf{Comm} & \textbf{RL inference} & \textbf{Total} \\
        \midrule
        CPU core assignment request & 1.00 & 2.57 & 2.57 \\
        Preemption time-slice update & 1.00 & 2.98 & 2.98 \\
        Scheduling overhead per task & 1.00 & 2.57 & 2.57 \\
        \bottomrule
    \end{tabular*}
    \label{tab:overhead_ratio}
    \vspace{-0.5cm}
\end{table}

The Pareto space provides a unified view of the latency--runtime trade-off. Among the configurations we measured, moving along the empirical frontier improves one objective only at the expense of the other. This characterization is particularly relevant to serverless environments, where applications exhibit diverse priorities: deadline-constrained workloads favor low-latency operating points, whereas cost-sensitive workloads prefer low-runtime ones.

\begin{figure}[b]
    \vspace{-0.4cm}
    \centering
    \includegraphics[width=0.85\linewidth]{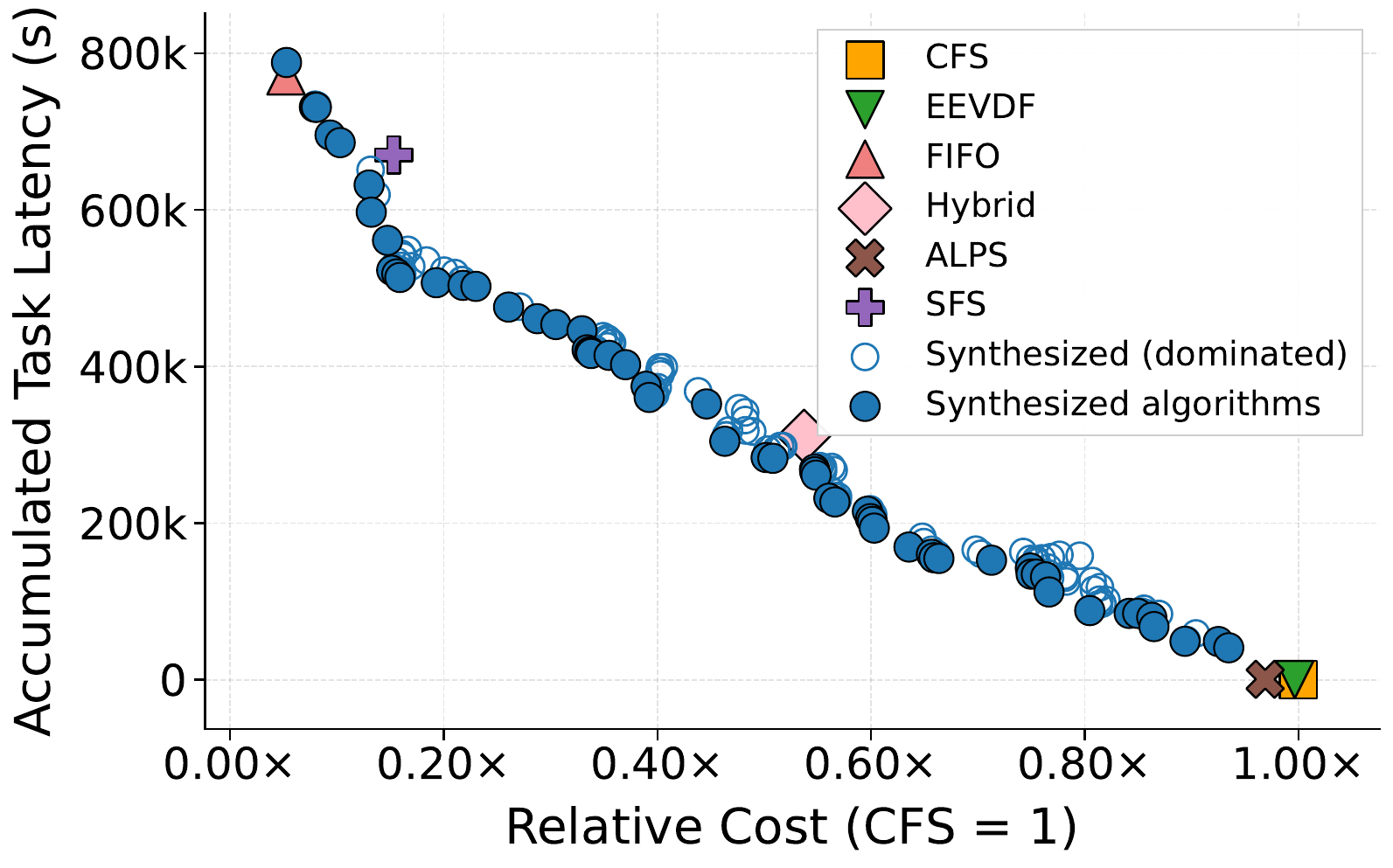}
    \vspace*{-0.4cm}
    \caption{Latency--cost trade-off. CFS, EEVDF, and ALPS~\cite{fu2024alps} achieve low latency with high cost, while FIFO and SFS~\cite{fu2022sfs} reduce cost but with higher latency. Hybrid~\cite{zhao2024serverless} provides intermediate baseline. Our synthesized schedulers span the trade-off space.}
    \label{fig:pareto_cost}
    \vspace*{-0.4cm}
\end{figure}

Figure~\ref{fig:pareto_cost} shows the latency--cost trade-off, where user cost is derived from accumulated task runtime. CFS lies in the low-latency, high-cost region, whereas FIFO reduces cost at the expense of higher accumulated task latency. By evaluating the synthesized scheduler under different preemption configurations, we obtain a set of operating points spanning these two extremes. A point is dominated if some other synthesized point is at least as good on both latency and cost, and better on one. We draw these as hollow markers; the solid points form the empirical Pareto frontier. These points trace a latency--cost trade-off curve spanning the region between the FIFO-like and CFS-like extremes, letting system operators pick an operating point that matches their latency targets and cost budgets in practical real-world FaaS scenarios.

\begin{figure}[t]
    \vspace{-0.3cm}
    \centering
    \includegraphics[width=0.8\linewidth]{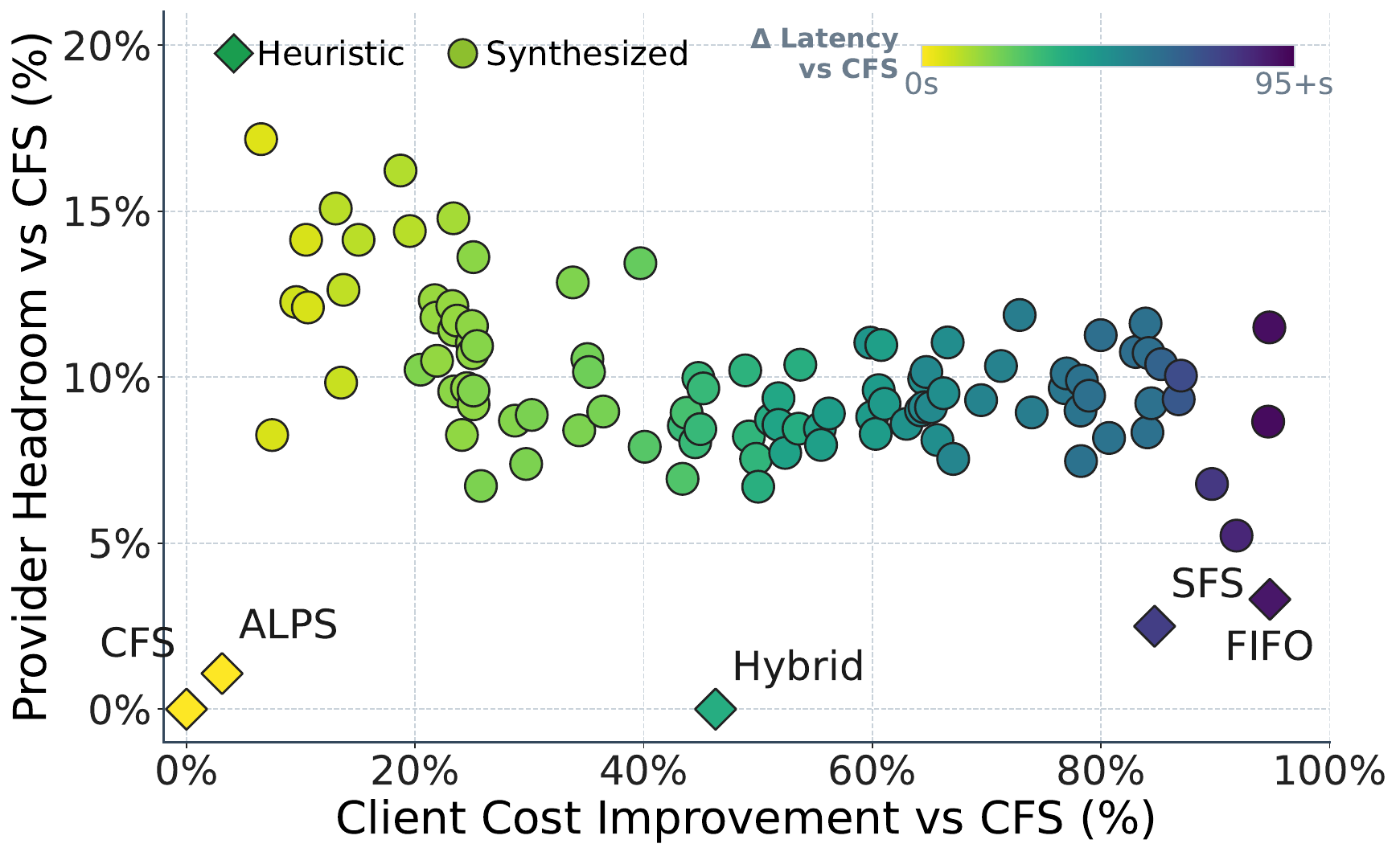}
    \vspace*{-0.35cm}
    \caption{Provider headroom versus client cost improvement, both relative to CFS. Synthesized schedulers can increase profit both for provider and client, depending on latency.}
    \vspace*{-0.5cm}
    \label{fig:provider_client}
\end{figure}

Figure~\ref{fig:provider_client} plots provider headroom against
client cost improvement, both measured relative to CFS. Provider headroom is defined by the amount of CPU time that the respective scheduling policy can clear compared to CFS. Effectively, provider headroom is a measure of how much more workload the cloud could pack on a single machine, increasing profit for the cloud provider. Each point in the figure is a policy evaluated on the same workload, the color indicates the mean per-task latency increase over CFS (light for low, dark for high). EEVDF and ALPS offer no improvements, neither for clients nor for providers. 

The state-of-the-art schedulers address a single objective: FIFO attains the largest cost improvement but at the highest latency and near-zero headroom, while Hybrid attains cost improvement but leaves no provider headroom. In contrast, the synthesized algorithms occupy the upper region, jointly achieving high headroom and cost improvement, with several of them keeping latency low. Overall, by exploring the empirical Pareto front of the trade-off space using synthesized schedulers we find interesting points which improve the profitability of both clients (barring increased latency), and cloud providers, allowing them to bin-pack more workloads onto the same resources.

\begin{figure}[t]
    \vspace{-0.3cm}
    \centering
    \includegraphics[width=0.85\linewidth]{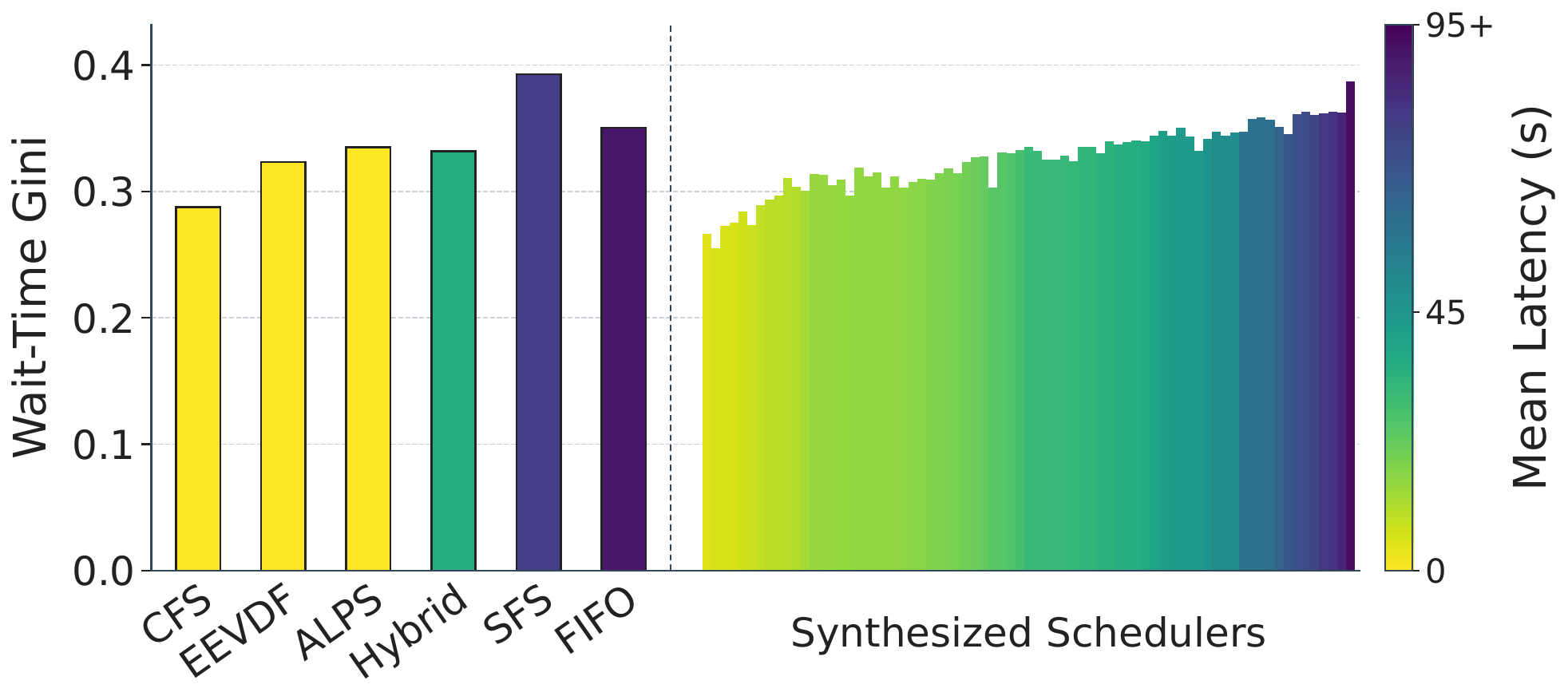}
    \vspace*{-0.4cm}
    \caption{Wait-time fairness and mean task latency across heuristic and synthesized schedulers according to the Gini coefficient. The synthesized schedulers offer similar fairness to existing schedulers.}
    \vspace*{-0.5cm}
    \label{fig:wait_gini}
\end{figure}

To ensure that the synthesized schedulers offer trade-offs in the latency--runtime space without unfairly starving tasks, we perform a Gini coefficient analysis~\cite{ogryczak2019fair}. Figure~\ref{fig:wait_gini} plots the wait-time Gini coefficient together with the mean task latency across all schedulers; a lower wait-time Gini indicates a more even distribution of waiting time across tasks, and the color represents mean task latency. The synthesized schedulers span a broad range of operating points, and their wait-time Gini stays within the same range as the heuristic schedulers. This shows that our synthesized schedulers offer fairness comparable to existing schedulers, and therefore do not obtain their latency--runtime trade-offs by starving tasks.

\vspace{-0.15cm}
\section{Related work}

\textbf{Learning-based resource management.}
Learning-based methods have been applied to a wide range of systems problems, including network congestion control~\cite{liao2024astraea, abbasloo2020classic,jay2019deep}, resource allocation and partitioning~\cite{yang2019miras, mao2016resource, qiu2020firm, mao2022mean, chen2021drlpart}, autoscaling~\cite{gari2021reinforcement, kardani2020adrl, schuler2021ai, veni2016auto, zafeiropoulos2022reinforcement}, and cluster scheduling~\cite{mao2019learning, zhang2022schedinspector, qiao2021pollux, jajoo2022slearn}. These demonstrate the value of learning and adaptation for systems decision making. However, they operate at the network, cluster, job, or resource-partitioning level, rather than at the OS scheduling layer of a serverless worker.

\textbf{Learning-based methods for serverless platforms.}
Recent work has also applied learning-based methods to serverless platforms~\cite{yu2021faasrank, qiu2022simppo, qiu2023aware, mampage2025deep}. These systems focus on platform-level function scheduling, resource provisioning, autoscaling, or hybrid serverless/serverful placement. In contrast, our work focuses on the OS-level scheduling on a worker node. We study how learning can synthesize OS-level scheduling policies for core assignment, preemption, and load rebalancing on a multi-core server.

\textbf{Serverless-aware OS-level scheduling.}
Several systems optimize serverless execution through OS-level scheduling mechanisms~\cite{kaffes2019centralized, kaffes2022hermod, abdi2023palette, fu2024alps, fu2022sfs, zhao2024serverless}. These systems improve over general-purpose schedulers by using mechanisms such as centralized dispatch, locality-aware placement, preemption, work stealing, or hybrid scheduling. Most of these approaches, however, rely on fixed heuristic policies. Our work uses a reinforcement learning architecture to synthesize schedulers that can be tuned toward different points in the latency--runtime trade-off space.

\textbf{Hardware- and data-aware serverless optimization.}
Beyond CPU scheduling policies, prior work has explored hardware- and data-aware serverless optimization, including memory-bandwidth and cache-aware scheduling~\cite{park2019copart, suresh2021servermore, gajanin2025performance}, local memory, container caching, and edge object-store mechanisms~\cite{li2022faasflow, yu2024rainbowcake, chen2025capybara}, Data Processing Units (DPUs)~\cite{liu2024fuyao}, and hardware-assisted preemption~\cite{li2024libpreemptible}. These techniques provide finer control over serverless execution resources and data locality. Our work focuses on a complementary direction: synthesizing tunable OS-level scheduling policies for CPU core assignment, preemption, and rebalancing.

\vspace{-0.15cm}
\section{Conclusion}
We presented a class of synthesized schedulers for the execution of serverless functions. Rather than relying on fixed heuristics, we synthesize scheduling policies using a reinforcement learning architecture that jointly learns core assignment, preemption, and rebalancing under different reward objectives. We implemented our schedulers on top of the Linux extensible scheduling framework \texttt{sched\_ext} and evaluated them on real-world Azure function traces. Our results show that synthesized schedulers are competitive with time-tested schedulers such as CFS, EEVDF, and FIFO, and can be tuned to a near-continuous set of operating points across the latency--runtime space, whereas each existing scheduler remains confined to a narrow region of the trade-off space. We show that synthesized schedulers can reduce user cost while improving platform profitability. More broadly, our results suggest that synthesizing schedulers is a practical path toward workload-specialized OS scheduling. We release the source code at: \textcolor{blue}{\url{https://github.com/ZhaoNeil/synthesized-scheduler.git}}

\vspace{-0.15cm}
\bibliographystyle{ACM-Reference-Format}
\bibliography{reference}

\end{document}